\documentclass[onecolumn, a4size, 11pt,final]{IEEEtran}
\usepackage{amsmath}
\usepackage{amssymb}
\usepackage{amsfonts}
\usepackage[final]{graphicx}
\usepackage{ctable}
\usepackage{algorithm}
\usepackage{algorithmic}
\usepackage{multirow}
 \usepackage{subfigure}

 \usepackage{cite}

\usepackage{subeqnarray}

\usepackage{amsmath,amsfonts,amssymb,epstopdf}

\usepackage{array}

\usepackage{bm}
\usepackage{mdwmath}
\usepackage{mdwtab}

\title{Green 5G Heterogeneous Networks through Dynamic Small-Cell Operation}

\author
{Shijie~Cai,
Yueling~Che,~\IEEEmembership{Member,~IEEE,}
Lingjie~Duan,~\IEEEmembership{Member,~IEEE,}
Jing~Wang,~\IEEEmembership{Member,~IEEE,}
Shidong~Zhou,~\IEEEmembership{Member,~IEEE,}
and~Rui~Zhang,~\IEEEmembership{Senior Member,~IEEE}
\thanks{This paper was presented in part as a poster at the 2nd IEEE Global Conference on Signal and Information Processing  (GlobalSIP).
This work is supported by the SUTD-ZJU Joint Collaboration Grant (Project Number: SUTD-ZJU/RES/03/2014). It is also supported by 
National Basic Research Program of China (2012CB316002),
Natural Science Foundation of China (61201192), the Science Fund for Creative Research Groups of NSFC (61321061),
863 project (2014AA01A704),
National S\&T Major Project (2014ZX03003003-002),
Key grant Project of Chinese Ministry of Education (No.313005),
the Open Research Fund of National Mobile Communications Research
Laboratory, Southeast University (2012D02),
Tsinghua-Qualcomm Joint Research Program, and Tsinghua-Intel International S\&T
Cooperation Program (ICRI-MNC).}
\thanks{S. Cai, J. Wang, and S. Zhou are with the Department of Electronic Engineering, Research Institute of Information Technology, Tsinghua National Laboratory for Information Science and Technology (TNList), Tsinghua University, Beijing 100084, China (e-mail: caisj06@gmail.com, \{wangj, zhousd\}@tsinghua.edu.cn).}
\thanks{Y. L. Che and L. Duan are with the Engineering Systems and Design Pillar, Singapore University of Technology and Design (e-mail: \{yueling\_che, lingjie\_duan\}@sutd.edu.sg). Y. L. Che is the corresponding author. }
\thanks{R. Zhang is with the Department of Electrical and Computer Engineering, National University of Singapore (e-mail: elezhang@nus.edu.sg).}
}

\begin{document}
\maketitle
\thispagestyle{empty}

\begin{abstract}
Traditional macro-cell networks are experiencing an upsurge of data traffic, and small-cells are deployed to help offload the traffic from macro-cells. Given the massive deployment of small-cells in a macro-cell, the aggregate power consumption of small-cells (though being low individually) can be larger than that of the macro-cell. Compared to the macro-cell base station (MBS) whose power consumption increases significantly with its traffic load, the power consumption of a small-cell base station (SBS) is relatively flat and independent of its load. To reduce the total power consumption of the heterogeneous networks (HetNets), we  dynamically change the operating states (on and off) of the SBSs, while keeping the MBS on to avoid  any service failure outside active small-cells. First, we consider that the wireless users are uniformly distributed in  the network, and propose an optimal location-based operation scheme by gradually turning off the SBSs closer to the MBS. We then extend the operation problem to a more general case where users are non-uniformly distributed in the network. Although this problem is NP-hard, we propose a joint location and user density based operation scheme to achieve near-optimum (with less than 1\% performance loss in our simulations) in polynomial time.
\end{abstract}

\begin{IEEEkeywords}
Green communication, heterogeneous networks (HetNets), small-cell operation, traffic offloading.
\end{IEEEkeywords}

\setlength{\baselineskip}{1.3\baselineskip}
\newtheorem{definition}{\underline{Definition}}[section]
\newtheorem{fact}{Fact}
\newtheorem{assumption}{Assumption}
\newtheorem{theorem}{\underline{Theorem}}%[section]
\newtheorem{lemma}{\underline{Lemma}}%[section]
\newtheorem{corollary}{\underline{Corollary}}%[section]
\newtheorem{proposition}{\underline{Proposition}}%[section]
\newtheorem{example}{\underline{Example}}[section]
\newtheorem{remark}{\underline{Remark}}%[section]
\newcommand{\mv}[1]{\mbox{\boldmath{$ #1 $}}}
\newtheorem{property}{\underline{Property}}[section]

\section{Introduction}
Driven by the exponentially increased wireless data traffic, the cellular network is expected to become increasingly heterogeneous to improve the spectral efficiency  as we move to the 5G cellular network \cite{Andrews.JSAC.14}.  However, the deployment of massive small-cells in the macro-cells  can   increase   the total power consumption of the 5G heterogeneous networks (HetNets) \cite{Hu.Mag.14}. The power consumption of a HetNet comes from both the macro-cell base stations (MBSs) and the small-cell base stations (SBSs). Researchers have paid much attention to study power saving at the MBS side, by dynamically changing the MBSs' on and off states to meet the stochastic traffic (see, e.g.,\cite{Wu2013,Soh2013}), but very few work has considered power saving at the SBSs.

As there are increasingly more  small-cells deployed in the 5G cellular network, their power consumption is not ignorable. Auer \emph{et al.} in \cite{D2.3} showed that in many European countries, the typical power consumption of a SBS is $10$W  and that of an MBS is $930$W. Thus, the power consumption of $100$ SBSs (or small-cells) is larger than that of an MBS (or macro-cell). As a result, it is important to jointly manage the power consumption of macro-cells and small-cells for more energy-efficient operation with traffic sharing. According to \cite{D2.3,Luo2013,Vereecken2011},   the MBS and  the SBS are different in their power consumption rates with their traffic loads. The MBS's power consumption increases exponentially with its traffic load in terms of the number of users served given each user has a constant rate requirement \cite{Luo2013}, while  the SBS's power consumption is  almost independent of its load and even flat for any load \cite{D2.3}, \cite{Vereecken2011}. Data offloading
from the MBS to the SBSs thus helps save the MBS's power consumption but inevitably requires more SBSs to be turned on and increases the  total  power consumption of the SBSs. The operator should be aware of this in designing the traffic sharing among the macro- and small-cells, and should also consider the heterogeneity of small-cells in location and user coverage.

This thus motivates this paper to investigate the dynamic adaptation of the SBSs' operation modes (on or off)  to save the total power consumption in the  5G HetNet while meeting  all users' service requirements. To study the SBSs' dynamic operation for the small-cells, we make a practical assumption  that the MBS is always active to provide the seamless coverage of control signal and avoid any service failure to serve users outside active small-cells\cite{xu2013}. Though we focus on small-cell dynamic operation, our approach does not exclude prior macro-cell operation schemes. The MBSs and  the SBSs are dynamically operated over different spatial and time scales. Small-cells are dynamically operated inside a macro-cell, while a macro-cell is  dynamically operated in a much larger range. Turning on or off an MBS usually takes several minutes\cite{HuaWei2012}, while a SBS can be quickly turned on or off in seconds\cite{Ashraf2011}.

The main contributions of this paper are summarized as follows:
\begin{itemize}
\item \emph{Novel small-cell dynamic operation to minimize the total HetNet power consumption:} We study dynamic small-cell on/off operation to serve offloaded traffic  from the macro-cell  for minimizing the total power consumption of the HetNet. We model   different power consumption patterns for both the MBS  and the SBSs with respect to their traffic load, where the SBSs'   locations and their user coverage areas are taken into account.
\item \emph{Small-cell dynamic operation to serve uniformly distributed users:} We take the spatial randomness of the user locations into consideration. We start with a special case with uniformly distributed users in the HetNet in Section III,  where user densities in all the small-cells and the macro-cell are identical.  We propose an optimal   location-based  operation algorithm to   decide the operation modes of the SBSs according to their distances to the MBS to minimize the total HetNet power consumption.
\item \emph{Small-cell dynamic operation to serve non-uniformly distributed users:} We then extend to a more general case with non-uniformly distributed users in the HetNet in Section IV, where the user density varies over different small-cells and the macro-cell. In this case, the HetNet power minimization problem is shown to be NP-hard. We propose a location-and-density-based operation algorithm that provides a near-optimal operation solution in  polynomial time to decide the SBSs' on/off states.
\item \emph{Performance evaluation:} In Section V, through extensive simulations, we show that the location-and-density-based operation algorithm achieves less than 1\% performance loss as compared to the optimal one. By comparing with two benchmark schemes with no SBS on/off adaptation and probability-based SBS on/off adaptation, respectively, we show that our proposed small-cell operation scheme can more efficiently save the total HetNet power consumption.  We also show that   the MBS's power consumption can even decrease with the increasing HetNet traffic load  under our proposed   scheme, as more small-cells are turned on to offload the macro-cell traffic.
\end{itemize}

Recently, HetNet dynamic operation for power saving draws significant attention   and many prior studies have focused on macro-cells' intelligent operations (e.g.,\cite{Niu2010, Weng2011, Marsan2009, Eunsung2013, Han2013, Bousia2012,Hossian.TWC.13,Che.JSAC.revised, Xu.JSAC.16}). Among these works, there are two major approaches: cell zooming and base station (BS) sleeping. The approach of cell zooming proposed in \cite{Niu2010} reduces the macro-cell power consumption by adjusting the cell size according to the covered traffic load, the quality-of-service (QoS) requirements and the channel conditions. The power saving potential of the cell zooming approach was further studied in \cite{Weng2011}. BS sleeping is the second major approach to save power by switching BSs between on and off. As a pioneering work, Marsan \emph{et al.} in \cite{Marsan2009} showed that 25-30\% of the total power consumption can be saved by reducing the number of active macro-cells when the traffic is low. Considering  the users' traffic variations over both space and time,  the authors in \cite{Che.JSAC.revised} jointly applied tools from stochastic geometry and dynamic programming to design the optimal BS on/off adaptation scheme.
The application and extension of this approach in practical systems have also been extensively studied in, e.g.,  \cite{Eunsung2013,Han2013,Bousia2012,Hossian.TWC.13, Xu.JSAC.16},  where various implementable BS  on/off
algorithms were proposed.   Besides these two approaches, user association for load balancing between macro-cells and small-cells was studied in \cite{Ye2013}.  However, all these studies  overlooked the impact of small-cells dynamic operation  on the total HetNet  power consumption.

As a preliminary work, Ashraf \emph{et al.} in \cite{Ashraf2011} showed that small-cell sleeping has great power-saving potential. The studies in \cite{Wu.15,Saker2012,Sinha2014} further exploited this potential and proposed small-cell control algorithms for power saving. However, these studies focused on small-cells only, and proposed to serve users within the small-cells. In practice, users in deactivated small-cells also need to be served by macro-cells. Thus, the different power consumption patterns for macro-cells and small-cells should be considered, and load balancing between them should be properly  designed  to reduce the total HetNet  power consumption. Though \cite{Cho2013} considered small-cells' activation, it did not aim to minimize the total HetNet power consumption but mitigate interference.

We summarize the key notations in this paper in Table I.

\begin{table}[htbp]
\tabcolsep 1.5mm \caption{Key notations}
\begin{center}
\begin{tabular}{|c|c|}
  \hline
  Notations&Descriptions\\
  \hline
  \hline
  $R_0$&Macro-cell radius\\
  \hline
  $R_s$&Small-cell radius\\
  \hline
  $\mathcal{M}=\{1,...,M\}$&Set of $M$ SBSs\\
  \hline
  $d_m$, $m\in \mathcal{M}$ & distance between SBS $m$ and the MBS \\
  \hline
  $K$&Number of macro-cell users \\
  \hline
  $r_k$, $k\in\{1,...,K\}$&Distance between macro-cell user $k$    and the MBS  \\
  \hline
  $r_0$&Reference distance in channel model\\
  \hline
  $D$&Reference path loss\\
  \hline
    $\alpha$&Path-loss exponent\\
    \hline
    $h_k$&Rayleigh fading between  user $k$ and the MBS\\
   \hline
  $A_m$&Coverage area of SBS $m$\\
  \hline
  ${A_0}$&  Area outside of all small-cells\\
  \hline
  $\bm{\theta}=\{\theta_1,...,\theta_M\}$&Operation modes of all $M$ SBSs \\
  \hline
  $\lambda_n$, $n\in\{0\}\cup \mathcal{M}$ &User density in $A_n$ \\
   \hline
  $P^t(\bm{\theta})$&MBS's transmit power to all macro-cell users \\
  \hline
  $P_{max}^t$&MBS's maximum transmit power \\
  \hline
  $P_k^t$&MBS's transmit power to user $k$ \\
  \hline
  $p_0$ &SBS's  power consumption  in sleeping mode   \\
   \hline
  $ p_1$&SBS's  power consumption  in  active mode  \\
  \hline
  $P^{Het}(\bm{\theta} )$&Total power consumption in the HetNet \\
  \hline
  $P_k^r$&Macro-cell user $k$'s received power from the MBS  \\
  \hline
 $W$ &Operation spectrum bandwidth of the macro-cell \\
  \hline
  $b$&each user's required data rate \\
  \hline
  $\varepsilon$&Maximum allowable outage probability\\
  \hline
\end{tabular}
\end{center}
\end{table}

\section{System Model}

\begin{figure}
\centering
\DeclareGraphicsExtensions{.eps,.mps,.pdf,.jpg,.png}
\DeclareGraphicsRule{*}{eps}{*}{}
\includegraphics[angle=0, width=0.7\textwidth]{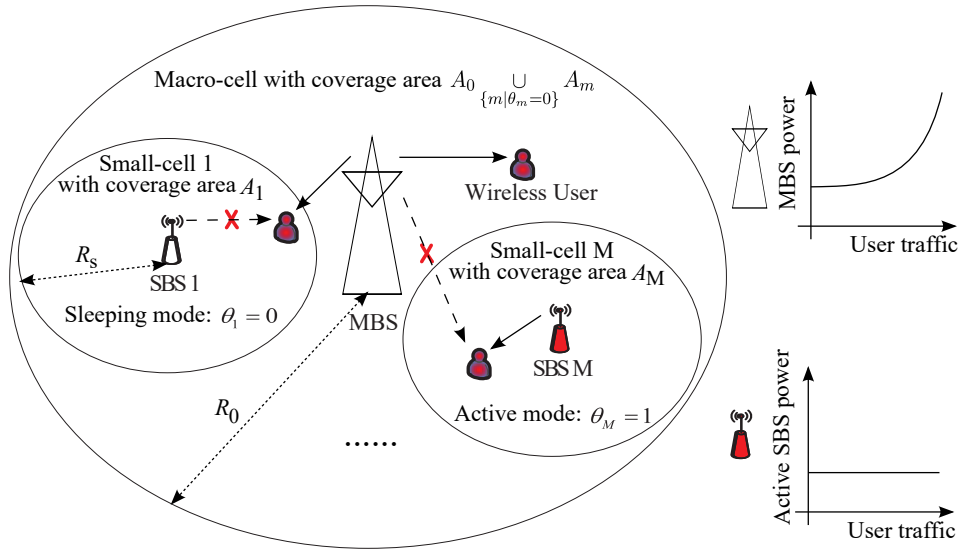}
\caption{The HetNet system model with one macro-cell and $M$ small-cells. }
\label{System_model}
%\vspace{-0.1in}
\end{figure}

As shown in Fig.~\ref{System_model}, we consider a two-tier HetNet, where $M$, $M\geq 1$, small-cells are deployed in a macro-cell. The $M$ SBSs dynamically adjust their on/off operations to serve the offloaded traffic from the MBS, so as to save the total power consumptions across all the SBSs and the MBS in the HetNet. In the following, we first present the network model of the considered HetNet. We then model the power consumption at each SBS and the MBS. At last, since the transmit power of the MBS increases over its served traffic load,  we   consider the impact of the traffic load served by the MBS and derive its transmit power.

\subsection{Network Model}

This subsection presents the network model. Denote the coverage radius of the MBS and each of the SBSs as $R_0$ and $R_s$, respectively, where $R_0> R_s>0$.
We assume the MBS and each of the SBSs are located at the center of their respective coverage areas.
Without loss of generality, we  assume the MBS is located at the origin $o$, given by $(0,0)$, of the two-dimensional plane $\mathbb{R}^2$.
Let $B(x,r)$, $x\in\mathbb{R}^2$, $r\in(0,\infty)$, denote a circle of radius $r$ centered at $x$. The area that can be covered by the MBS is thus given by $B(o,R_0)$.
The locations of all $M$ SBSs are given.
Denote the location of SBS $m$, $m\in \mathcal{M}\triangleq \{1,...,M\}$, as $x_m\in B(o,R_0)$, and the  coverage area of SBS $m$ as $A_m\triangleq B(x_m,R_s) \subset B(o,R_0)$.
It is easy to find that $A_m=\pi R_s^2$, $\forall m\in \mathcal{M}$.
We also denote the complement region of all small-cells' coverage areas in the macro-cell as ${A_0} \triangleq \overline {\mathop  \cup \limits_{m \in \,\mathcal{M}} {A_m}}\cap  B(o,R_0)$.

First, denote the operation mode of SBS $m\in\mathcal{M}$ by $\theta_m\in\{0,1\}$.
Each SBS  can  be either in the active mode (i.e., state $\theta_m=1$) to serve the data offloaded from the MBS or in the sleeping mode ($\theta_m=0$) to save its own power. The operation modes of all SBSs are given by a vector $\bm{\theta}=[\theta_1,\cdots,\theta_M]$.
The MBS is always active to provide the seamless control signal coverage and avoid any service failure\cite{xu2013}.
The coverage area of the MBS is thus  expressed as ${A_0}\mathop  \cup \limits_{\{ m|{\theta _m} = 0\} } A_m$.
To properly serve the users in the network, we adopt the ``separate carriers'' model such that the MBS   operates over a different spectrum band  from the SBSs, to avoid the interference between the MBS and the SBSs.{\footnote{This paper's focus is the HetNet power saving instead of spectrum allocation. A future extension may consider the other ``shared spectrum" model where macro-cells and small-cells  operate  over the same spectrum band. In this case, proper  interference control between the SBSs and the MBS as in \cite{Cho2013} is needed.}}   The practice of ``separate carriers'' has been widely applied in industry (e.g., by China Unicom\cite{Shetty2009,Duan2013,Wu1997}).
This paper  only focuses on one macro-cell, where the MBS efficiently exploits the available radio resources to support all the users in the macro-cell. As will be shown later in Section II-C,  the transmit power of the MBS is derived to assure the QoS of the users in the macro-cell.
Since all active SBSs  operate over the same spectrum band, the inter-cell interference between the active SBSs is generally unavoidable.
It is  noticed that   extensive schemes in the literature have been proposed to effectively control/mitigate the downlink inter-cell interference via efficient coordination between the small-cells (see e.g., \cite{Kosta.Survey.13} and the references therein).
As a result,  considering the short-transmission range in the small-cell as well as  the low transmit power levels of the SBSs, we adopt the following assumption  in this paper.
\begin{assumption}
In the considered HetNet, we assume the inter-cell interference between the active SBSs can  always be properly controlled  to assure the QoS of the small-cell users.
\end{assumption}

Next, we take the spatial randomness of the wireless users' locations into consideration, and apply the widely-used Poisson point processes (PPPs) to model the user locations in the HetNet (see, e.g.,  \cite{Cao2013}-\cite{Baccelli.NOW.I}). Specifically,  for each small-cell $m\in \mathcal{M}$, we use a homogeneous PPP with density $\lambda_m>0$ to model the locations of the  users in SBS $m$'s coverage area $A_m$. Similarly, we also use a homogeneous PPP with density $\lambda_0>0$ to model the locations of the users in the MBS's coverage area $A_0$. We assume all the PPPs are mutually independent. As a result, by the property of the PPP, the  users under this model are not only independently distributed over the entire HetNet, but also uniformly distributed with density $\lambda_m$ or $\lambda_0$ within each $A_m$, $m\in \mathcal{M}$, or $A_0$, respectively \cite{Baccelli.NOW.I}.
However, due to the non-identical user densities over $A_m$'s and $A_0$ in general, users' distributions are not uniform over different small-cells and the macro-cell.
This is reasonable as some small-cells are located as hotspots to serve the user crowds. We suppose that the users within the active small-cells are automatically served by the corresponding SBSs, while those outside all the active small-cells are served by the MBS. Such a control and access model allows us to turn off some SBSs to save energy, while keeping all the users being served.

In the following  subsection, we introduce the power consumption models for the MBS and the SBSs, and derive the   total  power consumption across all SBSs and  the MBS in the HetNet.

\subsection{Power Consumption Models}
This subsection introduces the power consumption models for the MBS and the SBSs. We first consider the power consumption model for the MBS.
The power consumption $P$ of the MBS starts from a base level $\underline{P}>0$ and increases linearly with its downlink transmit power $P^t$\cite{D2.3}. As $P^t$ increases with the macro-cell traffic after offloading to active small-cells, $P^t$ is a function of the SBSs' operation modes $\bm{\theta}$. We thus rewrite $P^t$ as $P^t(\bm{\theta})$ and have
 \begin{equation}
\label{P_macro}
P = \underline{P} + uP^t(\bm{\theta}),
\end{equation}
where $u>0$ is the power utilization coefficient for the MBS. Take Europe for example, typically we have $\underline{P}=712$W and $u=14.5$\cite{D2.3}.

Next, consider the power consumption model for each SBS. According to \cite{D2.3} and\cite{Vereecken2011}, depending on its operation mode, we model the power consumption of  SBS $m$, $m\in \mathcal{M}$, as follows:
\begin{align}
\label{p_small}
p_m = \left\{ {\begin{array}{*{20}{l}}
{p_1=\underline{p}+vp^t,}&{\text{if}\;\theta_m=1\;\text{(active/on),}}\\
{p_0,}&{\text{if}\;\theta_m=0\;\text{(sleeping/off),}}
\end{array}} \right.
\end{align}
where $p_1$ and $p_0$ are the total power consumption of SBS $m$ in the active and sleeping modes, respectively, $\underline{p}>0$,  $p^t>0$, and $v>0$ are the base power level, the transmit power level, and the power utilization coefficient of the  SBS $m$ when it is in the active mode, respectively. It is noted that different from the MBS that consumes most power in power amplifier,   for a SBS, the load-dependent power amplifier is no longer the main power-consuming component \cite{Debaillie}, and its transmit power $p^t$ increases mildly with the traffic load (e.g., only $0.07$W increase as traffic load increases from 80\% to 100\%), due to the short-range communication in the small-cell \cite{D2.3}, \cite{Vereecken2011}. Therefore, we assume  $p^t$ and thus $p_1$  in (\ref{p_small}) are not related to the SBS's traffic load and are both constants for simplicity.  From \cite{D2.3}, the  typical values of $p_1$ and $p_0$ are $10$W and $3$W, respectively.
By comparing \eqref{P_macro} and \eqref{p_small},  it is easy to find  that in a practical macro-cell  that can include hundreds or even thousands of small-cells, the power consumption of just $100$ active SBSs becomes larger than that of an MBS, which motivates our proposed    HetNet power saving via the SBS on/off adaptations.

At last, we give the overall power consumption in the HetNet based on \eqref{P_macro} and \eqref{p_small}. Given $\bm{\theta}$, we define $H(\bm{\theta} )\triangleq \sum_{m\in M} \theta_m$ as the number of active small-cells. Let $\Delta p=p_1 - p_0$. By summing  the power consumption over all the active SBSs and the MBS, the total power consumption in the HetNet is obtained as
\begin{equation}
\label{p_network}
P^{Het}(\bm{\theta} ) = \underline{P} + u{P^t(\bm{\theta} )} + M{p_0} + H(\bm{\theta} )\Delta p.
\end{equation}

Since the MBS's transmit power $P^t(\bm{\theta} )$ is determined by its served traffic load as well as the operation modes of the SBSs, to more explicitly express $P^{Het}(\bm{\theta} ) $ in (\ref{p_network}), we derive $P^t(\bm{\theta} )$ in the following subsection.

\subsection{Derivation of MBS Transmit Power $P^t(\bm{\theta})$}

Based on the PPP-based user location model introduced in Section II-A, in this subsection we derive  the MBS's transmit power $P^t(\bm{\theta})$ for a given  operation mode  $\bm {\theta}$ of the SBSs. We first study the MBS's transmit power for each individual user in its coverage area, so as to satisfy the user's QoS requirement. Then by aggregating the MBS's transmit power for each user, we obtain the MBS's transmit power $P^t(\bm{\theta})$.
\subsubsection{MBS Transmit Power to Each Individual User}
We first focus on the MBS's transmit power for each individual macro-cell user that is located in its coverage area ${A_0}\mathop  \cup \limits_{\{ m|{\theta _m} = 0\} } A_m$. For a given operation mode  $\bm {\theta}$ of the SBSs, denote the number of macro-cell users  as $K$, and the   distance between the MBS and each macro-cell user  $k\in\{1,...,K\}$  as $r_k$.
We consider both  distance-dependent path loss and  short-term Rayleigh fading for the wireless channel between the MBS and each of its served users.
If the distance $r_k$ is shorter than a reference distance $r_0>0$, the MBS's transmit power to macro-cell user $k$, denoted by $P_k^t$, experiences a fixed path loss $D>0$. Otherwise, it attenuates with the distance $r_k$ according to the path-loss exponent $\alpha>0$. We also denote the Rayleigh fading channel from the MBS to macro-cell user   $k$ as $h_k$, which follows exponential distribution with unit mean.  We assume $h_k$'s are mutually independent. If the MBS transmits with power $P_k^t$ to user $k$, the received power, denoted by $P_k^r$, is then obtained as
\begin{align}
\label{channel_model}
P_k^r = \left\{ {\begin{array}{*{20}{l}}
{P_k^t{h_k}D{{\left(\frac{{{r_k}}}{{{r_0}}}\right)}^{ - \alpha }},} &{ \text{if}\;{r_k} \ge {r_0},}\\
{P_k^t{h_k}D,} & {\text{ otherwise.}}
\end{array}} \right.
\end{align}

We say the QoS requirement of a macro-cell user  $k$ is satisfied if the outage probability that the achieved data rate of  user $k$ being  smaller than $b$ bits/sec is no larger than a given threshold  ${\varepsilon}\ll 1$.
To find the transmit power $P_k^t$ that can assure the QoS of each macro-cell user, we adopt the widely-used equal-bandwidth-sharing  scheme among the users (see, e.g.,\cite{Cao2013,Singh2013,{Luo2013}}). Denote  the operated spectrum bandwidth of the MBS   as $W$. The assigned bandwidth for each user is thus $W/K$.{\footnote{To focus on  the transmit power derivation, we  consider the  equal-bandwidth-sharing  scheme for simplicity. The analysis method in this paper can also be applied to other channel allocation schemes (e.g., the unequal-bandwidth-sharing scheme in \cite{Guo2014}) in a similar manner. }} Similar to the transmit power derivation in \cite{Luo2013},  by using the received power $P_k^r$ and the bandwidth $W/K$ to  calculate the achieved data rate of each user $k$ based on the Shannon's formula, we can first find  the minimum required received power for user $k$ such that the achieved data rate is equal to $b$. Then by noticing that $h_k$ follows exponential distribution with unit mean, it is  easy to find the probability that
$P_k^r$
being less than the calculated minimum required received power is no larger than the threshold  ${\varepsilon}$, from which we  inversely calculate the   transmit power $P_k^t$  to assure macro-cell user $k$'s QoS  and obtain
\begin{equation}
\label{P_n_t}
P_k^t = \left\{ {\begin{array}{*{20}{l}}
{\frac{{\Gamma {N_0}W}}{{ - D\ln (1 - {\varepsilon})}} \times  \frac{{{2^{\frac{Kb}{W}}} - 1}}{K} \times {{\left(\frac{{{r_k}}}{{{r_0}}}\right)}^\alpha },}&{\text{if}\;{r_k} \ge {r_0},}\\
{\frac{{\Gamma {N_0}W}}{{ - D\ln (1 - {\varepsilon})}} \times \frac{{{2^{\frac{Kb}{W}}} - 1}}{K},}&{\text{otherwise},}
\end{array}} \right.
\end{equation}
where $N_0$ is the noise power density and $\Gamma \geq 1$ accounts for the loss of capacity due to practical coding and modulation. From \eqref{P_n_t},  the transmit power $P_t^k$ increases with the distance $r_k$ and the number of users $K$. Thus, we rewrite $P_k^t$ as $P_t^k(r_k,K)$, i.e., a function of $r_k$ and $K$.

 \newcounter{TempEqCnt}                         % 创建临时变量TempEqCnt
\setcounter{TempEqCnt}{\value{equation}} % 将当前公式序号 赋给TempEqCnt
\setcounter{equation}{7}
\begin{figure*}
\begin{equation}
\label{T1}
T(\bm{\theta} ) = \frac{{\Gamma {N_0}W}}{{ - D\ln (1 - \varepsilon )}}   \left[\exp \left[\left(2^{\frac{b}{W}} - 1\right)\left({\lambda _0}\pi R_0^2- {\lambda _0}M\pi R_s^2 + \sum\limits_{\{ m|{\theta _m} = 0\} } \lambda _m \pi R_s^2\right) \right] - 1\right].
\end{equation}

\begin{equation}
\label{C1}
Z(\bm{\theta} ) = \frac{\frac{2\pi \lambda _0}{\alpha  + 2}\left(R_0^{\alpha  + 2} + \frac{\alpha r_0^{\alpha  + 2}}{2}\right) - {\lambda _0}\sum\limits_{m = 1}^M {\pi R_s^2d_m^\alpha }+\sum\limits_{\{ m|{\theta _m} = 0\} } {{\lambda _m}\pi R_s^2d_m^\alpha } }{r_0^\alpha \left(\lambda _0\pi R_0^2 - \lambda _0 M\pi R_s^2 + \sum\limits_{\{ m|{\theta _m} = 0\} } \lambda _m \pi R_s^2\right)}.
\end{equation}
\hrulefill
\end{figure*}
 \setcounter{equation}{\value{TempEqCnt}}

\subsubsection{MBS Transmit Power $P^t(\bm{\theta})$ to All Users}
We now derive the MBS's   total transmit power  $P^t(\bm{\theta})$  for all the macro-cell users in its coverage area. According to the PPP-based user location model in Section II-A, it is noticed  that the macro-cell user number $K$ and the distance $r_k$ between the macro-cell user $k$ and the MBS   which determine $P_t^k(r_k,K)$ in \eqref{P_n_t} are both random variables. It is easy to obtain that $K$ is a Poisson distributed random variable with mean $\mu$ being equal to  the average number of users in the MBS's coverage area ${A_0}\mathop  \cup \limits_{\{ m|{\theta _m} = 0\} } A_m$. We thus have  $\mu=\lambda_0\|A_0\|+\sum_{\{m|\theta_m=0\}}\lambda_m \|A_m\|$, where $\|A_n\|$ denotes the area of $A_n$, $\forall n\in\{0\}\cup \mathcal{M}$.
Moreover,    given the macro-cell user number $K$, all the macro-cell users are identically and independently distributed in the MBS's coverage area \cite{Baccelli.NOW.I}. As a result,    $r_k$'s, $\forall k=\{1,...,K\}$, are identical  and  independently distributed (i.i.d.) random variables.
Therefore, for a given operation mode $\bm{\theta}$ of all the SBSs, by summing $P_t^k(r_k,K)$ over all users in the MBS's coverage area and taking expectations over $K$ and each $r_k$, we obtain    $P^t(\bm{\theta})$ as
\setcounter{equation}{5}
\begin{align}
\label{P_t_avg}
{P^t}(\bm{\theta} ) &= \mathtt{E}_K\left[{{\mathtt{E}}_{r_1,\cdots,r_K}}\left[\sum\limits_{k = 1}^K {P_k^t({r_k},K)}\right ]\right] \nonumber \\
&\overset{(a)}{=}{{\mathtt{E}}_K}\left[K   \mathtt{E}_{r_k}\left[P_k^t({r_k},K)\right]\right],
\end{align}
where  equality $(a)$ follows since $r_k$'s are i.i.d. random variables for a given $K$.
In the following, by first deriving the inner expectation  $\mathtt{E}_{r_k}\left[P_k^t({r_k},K)\right]$ of (\ref{P_t_avg}) for a given $K$,   and   then   deriving  the outer expectation of  (\ref{P_t_avg}) by using the fact that $K$ follows Poisson distribution with mean $\mu$, we obtain an explicit expression of ${P^t}(\bm{\theta} )$ in Theorem~1.

\begin{theorem}
 Given the operation mode  $\bm{\theta}$ of all $M$ SBSs, the transmit power ${P^t}(\bm{\theta} ) $ of the MBS  is the product of a  \emph{macro-cell's traffic factor} $T(\bm{\theta} )$ and a  \emph{power efficiency factor} $Z(\bm{\theta} )$, i.e.,
 \begin{equation}
\label{Pt_macro_avg}
{P^t}(\bm{\theta} ) =T(\bm{\theta} )  Z(\bm{\theta}),
\end{equation}
where $T(\bm{\theta} )$  and $Z(\bm{\theta})$ are given in (\ref{T1}) and (\ref{C1}), respectively, and
 $d_m=|x_m|$  in (\ref{C1}) is the distance between  SBS $m$ and the MBS.
\end{theorem}

\begin{IEEEproof}
 Please refer to Appendix A.
\end{IEEEproof}

 \begin{remark}
The power efficiency factor $Z(\bm{\theta} )$ gives the MBS's average transmit power for an individual user, and the traffic factor $T(\bm{\theta} )$ gives the total   traffic load in the area  outside of all the active small-cells.
It is observed from $T(\bm{\theta})$ in \eqref{T1} that  similar to \cite{Luo2013}, the MBS's transmit power $P^t(\bm{\theta})$ in \eqref{Pt_macro_avg}   increases exponentially with its traffic load for assuring each macro-cell user's QoS.
Thus, we can turn on more SBSs  to offload the MBS's heavy traffic load for saving the MBS's power consumption. However, a large number of active SBSs also consume high power. As a result,  by adapting the operation modes of all the SBSs, an optimal trade-off between saving the MBS's power consumption and saving the SBSs' power consumption needs to be decided for minimizing the total HetNet power consumption.
\end{remark}

In the following two  sections, to properly study the HetNet power minimization problem, we start with a special case with uniformly distributed users in the HetNet where $\lambda_0=\lambda_1=\cdots=\lambda_M$. We then extend the results to a more practical case with non-uniformly distributed users where $\lambda_0$ and $\lambda_m$'s are not identical in general.

\section{Small-Cell Dynamic Operation for Uniformly Distributed Users}
This section studies the case where the users are uniformly distributed in the HetNet with identical user density, i.e., ${\lambda _0} = {\lambda _1} =  \cdots  = {\lambda _M}$.
In the following, we first formulate the HetNet power minimization problem in this case. We then provide a tractable method to solve this problem. At last, based on the optimal SBSs' operation mode, we study the impact of user density in deciding the SBSs' on/off operations.

\subsection{ Problem Formulation for Uniformly Distributed Users}
This subsection formulates the HetNet power minimization  problem via the dynamic on/off operations of the $M$ SBSs in the case with uniformly distributed users. The problem objective is to minimize the total HetNet power consumption over the MBS and all $M$ SBSs, which is defined in \eqref{p_network}. To obtain the expression of the total HetNet power consumption in the case with uniformly distributed users, we first substitute ${\lambda _0} \! = \!{\lambda _1} \!=  \!\cdots  \!= \!{\lambda _M}$ into \eqref{T1} and \eqref{C1},  and find $T(\bm{\theta} )$ and $Z(\bm{\theta} )$ in the uniformly distributed user case are simplified as
\setcounter{equation}{9}
\begin{equation}
\label{T2}
T(\bm{\theta} ) = \frac{\Gamma {N_0}W \! \left[\exp \! \left(\left(2^{\frac{b}{W}}\! -\! 1\right)\!\lambda_0\pi \! \left(R_0^2 \!- \! H(\bm{\theta} )  R_s^2\right)\right) \!- \!1\right]}{ - D\ln (1 - \varepsilon)}
\end{equation}
and
\begin{equation}
\label{C2}
Z(\bm{\theta} )= \frac{{\frac{{2\pi }}{{\alpha  + 2}}\left(R_0^{\alpha  + 2} + \frac{{\alpha r_0^{\alpha  + 2}}}{2}\right) - \pi R_s^2  \sum\limits_{\{ m|{\theta _m} = 1\} } {d_m^\alpha } }}{{r_0^\alpha   \pi \big(R_0^2 - H(\bm{\theta} )   R_s^2\big)}},
\end{equation}
respectively. Then by substituting \eqref{T2} and \eqref{C2} into \eqref{Pt_macro_avg}, we obtain ${P^t}(\bm{\theta} )$ in the  uniformly distributed user case. By substituting ${P^t}(\bm{\theta} )$ in this case into \eqref{p_network}, one can easily find  the total HetNet power consumption in   the  uniformly distributed user case. Moreover, by  practically considering that the transmit power $P^t(\bm{\theta})$ of the MBS  is constrained by a maximum allowable transmit power level \cite{D2.3}, which  is denoted by   $P_{max}^t$, we   formulate the HetNet power minimization problem as
\begin{align*}
({\mathop{\rm P}\nolimits} 1):\mathop {\min }\limits_{\bm{\theta}}  \;\;&P^{Het}(\bm{\theta})=\underline P  + uP^t({\bm{\theta}} )+ M{p_0} + H({\bm{\theta}} )\Delta p,\nonumber\\
s.t.\;\;&T({\bm{\theta}} )   Z({\bm{\theta}} ) \le P_{max }^t,\nonumber\\
&{\theta _m} \in \{ 0,\;1\} ,\;m \in \mathcal{M},
\end{align*}
where $T(\bm{\theta} )$ and $Z(\bm{\theta} )$ are given in (\ref{T2}) and (\ref{C2}), respectively.

It is noticed that if the traffic load in the HetNet is unexpectedly heavy, such that even if all the SBSs are active with  $\bm{\theta}=[1,...,1]$ to offload the traffic from the MBS, the MBS's required transmit power $P^t(\bm{\theta})=T({\bm{\theta}} )  Z({\bm{\theta}} )$ still exceeds $P_{max }^t$, problem (P1) becomes infeasible.
To avoid the trivial case without any feasible solution for the SBSs' operation modes $\bm{\theta}$ in (P1) for the uniform user distribution case, we assume the   traffic load in the HetNet always satisfies $T({\bm{\theta}} )  Z({\bm{\theta}} )|_{\bm{\theta}=[1,...,1]}\le P_{max }^t$.  As will be shown later in Section IV,  we also assume the system is always feasible when solving the HetNet power minimization problem  for the non-uniform user distribution case.
For the infeasible case  with unexpectedly heavy traffic load such that $T({\bm{\theta}} )   Z({\bm{\theta}} )|_{\bm{\theta}=[1,...,1]}> P_{max }^t$, one can apply admission control to allow only a portion of the users that satisfy $T({\bm{\theta}} )   Z({\bm{\theta}} )|_{\bm{\theta}=[1,...,1]}\le P_{max }^t$ to access the HetNet. However, the detailed admission control is beyond the scope of this paper.   A numerical example to show the HetNet total power consumption performance under the infeasible case is  given  in Section V-B.

Due to the binary operation mode of  each MBS, there are in total   $2^M$ combination possibilities  that need to be searched to find the optimal solution $\bm{\theta}^*=[\theta_1^*, \cdots,\theta_M^*]$ to problem (P1). To avoid such an exhaustive searching method with exponentially increased computational complexity over $M$,  we propose a  more efficient method   to solve (P1) in the following subsection by gradually turning off the SBSs according to their distances to the MBS.

\subsection{Optimal  Operation Modes of SBSs}
This subsection studies the optimal solution  to problem (P1). It is observed from (\ref{P_n_t}) that the transmit power of the MBS for each macro-cell user increases exponentially over the  distance between them. Thus, with identical  traffic load on average for  each  small-cells   in the case with uniformly distributed users,  the distance  between the SBS  and the MBS becomes an important criterion to decide which SBS should be turned off and let the  MBS  serve the users located in this small-cell. As a result, in the following, we present a SBSs' location-based policy to solve (P1), which shows   the SBSs that are close to the MBS   have a high  priority to be turned off.

\begin{proposition}\label{proposition: 1}
In the case with  uniformly distributed users, under the optimal operation mode  $\bm{\theta}^*=[\theta_1^*, \cdots,\theta_M^*]$ to problem (P1), if SBS $m \in \mathcal{M}$ is turned off with $\theta_m^*=0$,  any other SBS $n \in \mathcal{M}$ that is closer to the MBS than SBS $m$, i.e., $d_n<d_m$, is also turned off  with $\theta_n^*=0$.
\end{proposition}

 \begin{IEEEproof}
  Please refer to Appendix B.
 \end{IEEEproof}

 \begin{remark}
Proposition \ref{proposition: 1} implies that   since   the MBS consumes less power to serve the users that are located in the  close small-cells from (\ref{P_n_t}),  the SBS  that is close to the MBS has a higher priority  to be  turned off as compared to that  far from the MBS, so as to save the HetNet total power consumption.
We thus reorder  the SBSs in $\mathcal{M}$ according to their individual distances to the MBS, and obtain a new \emph{location-based set} $\mathcal{M}'=\{1,...,M\}$, in which for any two SBSs with indexes $m$ and $n$, respectively, we have $d_m\leq d_n$ if $m<n$.  Clearly, by assuming all SBSs are initially active, Proposition \ref{proposition: 1} implies that one can gradually turn off the SBSs according to the index order in  the location-based set $\mathcal{M}'$.
 \end{remark}

It is also noted that    turning off a SBS $m$ can not only decrease the power consumption of the SBS~$m$ by $\Delta p=p_1-p_0$, but also increase  the MBS's power consumption, where  the increased   power consumption at the MBS  is given by
 \begin{align}
\label{delta_p_m}
\Delta P_m= uP^t([\bm{0}_{1\times (m-1)}, \theta_m, \bm{1}_{1 \times (M-m)}])|_{{\theta _m} = 0}-uP^t([\bm{0}_{1\times (m-1)}, \theta_m ,\bm{1}_{1 \times (M-m)}])|_{{\theta _m} = 1}.
\end{align}
By studying $\Delta P_m$ over $m$, we obtain the following lemma.

\begin{lemma}
 By turning off the SBSs according to their index order in the location-based set $ \mathcal{M}'$,  the increased power consumption  $\Delta P_m$ at the MBS due to turning off SBS $m$  increases over $m \in \mathcal{M}'$.
\end{lemma}

 \begin{IEEEproof}
  Please refer to Appendix C.
 \end{IEEEproof}

As a result, according to Lemma 1, in the process of gradually deactivating the SBSs, while the  decreased power consumption at each deactivated SBS is a constant $\Delta p$, the MBS' power consumption increases each time for tuning off a SBS gradually. As a result, to save the  HetNet total power consumption over all SBSs and the MBS, the gradual deactivating process for the SBSs needs to stop at SBS $m$ once there is no power saving benefit at SBS $m+1$ (i.e., $\Delta p - \Delta P_{m+1} \le 0$).   Besides, since the MBS's transmit power cannot exceed the maximum transmit power $P_{max}^t$, the gradual deactivating process for the SBS also needs to stop once $P_{max}^t$ is reached. As a result, based on Proposition 1 and Lemma 1, we obtain a location-based algorithm to decide the operation modes of all the SBSs, which is given by Algorithm \ref{algorithm_uniform}.  In Algorithm \ref{algorithm_uniform}, we  define two integer thresholds $m_{th}^1$ and $m_{th}^2$ as
\begin{equation}
\label{m_th_1}
m_{th}^1 = \left\{ {\begin{array}{*{20}{l}}
{0,}&{\text{if}\,\Delta {P_1} \ge \Delta p,}\\
{m,}&{\text{if}\,\Delta {P_m} < \Delta p,\,\Delta {P_{m + 1}} \ge \Delta p,}\\
{M,}&{\text{if}\,\Delta {P_M} < \Delta p}
\end{array}} \right.
\end{equation}
and
\begin{equation}
\label{m_th_2}
m_{th}^2 = \left\{ {\begin{array}{*{20}{l}}
\!\!{0,}&\!\!\!\!{\text{if}\,P^t([{\theta _1} = 0,{\bm{1}_{1 \times (M - 1)}}])> P_{max }^t,}\\
\!\!{m,}&\!\!\!\!{\text{if}\,P^t([{\bm{0}_{1\! \times\! (m - 1)}},{\theta _m}\! = \!0,{\bm{1}_{1\! \times\! (M - m)}}])\!\le \! P_{max }^t,}\\&
\!\!{\quad \!\!\! P^t([{\bm{0}_{1\! \times m}},{\theta _{m  +  1}}\! = \!0,{\bm{1}_{1 \!\times\! (M - m - 1)}}])\!>\! P_{max }^t,}\\
\!\!{M,}&\!\!\!\!{\text{if}\,P^t([{\bm{0}_{1 \times M}}]) \le P_{max }^t,}
\end{array}} \right.
\end{equation}
respectively.

\begin{algorithm}[t]
\caption{Location-Based Operation Algorithm for SBSs in the Case with Uniformly Distributed Users}
\label{algorithm_uniform}
\begin{small}
\begin{algorithmic}[1]
\STATE Reorder all the small-cells in a new set $\mathcal{M}'$ as $d_m$'s increase
\STATE $L \leftarrow \pi R_a^2 - M\pi R_s^2$
\STATE $LR \leftarrow \frac{{2\pi }}{{\alpha  + 2}}\left(R_0^{\alpha  + 2} + \frac{{\alpha r_0^{\alpha  + 2}}}{2}\right) - \pi R_s^2\sum\limits_{m = 1}^{{M}} {d_m^\alpha }$
\STATE ${Q_1} = \frac{{\Gamma {N_0}W}}{{- D\ln (1 - \varepsilon )}}$
\STATE${Q_2} = {2^{\frac{b}{W}}} - 1$
\STATE ${P ^t} \leftarrow {Q_1} \times\frac{{LR}}{{r_0^\alpha   L}} \times \left({e^{{Q_2} {\lambda _0}  L}} - 1\right)$
\STATE ${P^{Het}} \leftarrow \underline{P} + u{P^t} + Mp_1$
\FOR{$m = 1:M$}
\STATE $L \leftarrow L + \pi R_s^2$
\STATE $LR \leftarrow LR + \pi R_s^2d_m^\alpha$
\STATE $P^{t'} \leftarrow {Q_1} \times \frac{{LR}}{{r_0^\alpha    L}} \times \left({e^{{Q_2}    {\lambda _0}   L}} - 1\right)$
\STATE $\Delta P_m~\text{in}~\eqref{delta_p_m}  \leftarrow u    P ^{t'} - u   { P ^t}$
\IF{$P ^{t'} \!> \!P_{max}^t~\!(\textrm{i.e.,~}m\!-\!1\!=\!m_{th}^2~\text{in}~\eqref{m_th_2})$ or $\Delta P_m \!\ge \! \Delta p~(\textrm{i.e.,~}m-1=m_{th}^1~\text{in}~\eqref{m_th_1})$}
\STATE RETURN $P^{Het}$, BREAK
\ENDIF
\STATE ${P^{Het}} \leftarrow {P^{Het}}+\Delta P_m-\Delta p$
\STATE ${P^t} \leftarrow P^{t'}$
\ENDFOR
\STATE RETURN $P^{Het}$
\end{algorithmic}
\end{small}
\end{algorithm}

\begin{theorem} \label{theorem: 2}
 Algorithm~\ref{algorithm_uniform} gives the SBSs' optimal operation mode $\bm{\theta}$ for problem (P1). In Algorithm~\ref{algorithm_uniform},  all the SBSs are initially set as  active and reordered based on their individual distances  to the MBS from the shortest to the largest in $\mathcal{M}'$, and then  are gradually turned off  based on their index order in $\mathcal{M}'$  until reaching SBS $m_{th}=\min\left(m_{th}^1, m_{th}^2\right)$, with  $m_{th}^1$ and $m_{th}^2$  given by (\ref{m_th_1}) and (\ref{m_th_2}), respectively.
\end{theorem}

\begin{IEEEproof}
 Please refer to Appendix D.
\end{IEEEproof}

It is easy to find that given the  location-based set $\mathcal{M}'$,  the computational complexity of Algorithm \ref{algorithm_uniform} increases linearly over $m_{th}$.
It is also observed from \eqref{m_th_2} that in Algorithm 1,   the operation threshold $m_{th}^2$ is  increasing over $P_{max}^t$, since more users can be supported by the MBS with an increased $P_{max}^t$.
However, it is not easy to obtain the varying trend of the threshold $m_{th}^1$ from \eqref{m_th_1}. In the following subsection, by focusing on the impact of  macro-cell user density, we study the threshold $m_{th}^1$ in detail.

\subsection{Impact of User Density in Small-Cell Operation}

In this subsection,  to focus on studying the threshold $m_{th}^1$, we  assume  there is no transmit power constraint at the MBS, i.e., $P^t_{max}=\infty$.
In this case, from \eqref{m_th_2}, it is easy to find that  $m_{th}=\min\left(m_{th}^1, m_{th}^2\right)=m_{th}^1$.
In the following, by defining the following function
\begin{equation}
\label{C_fun}
f(x)=\frac{{u\Gamma {N_0}W}}{{ - D\ln (1 - \varepsilon ) r_0^\alpha}} \times \frac{\exp\left(\left(2^{\frac{b}{W}} - 1\right)x\right) - 1}{x},
\end{equation}
we derive two  thresholds for the macro-user density $\lambda_0$, which  can  decide the value of $m_{th}^1$  in \eqref{m_th_1} and thus the SBSs' operation  modes.

\begin{proposition}\label{proposition: 2}
Under the setup of uniformly distributed users and assuming $P^t_{max}=\infty$, there always exists a user-density threshold $\lambda _{\mathtt{th}}^{\mathtt{off}}$, which is the unique solution to
\begin{align}
\label{lamda_th_1_uniform}
\Delta p =&\frac{{2 \lambda _{\mathtt{th}}^{\mathtt{off}} \pi}}{{\alpha  + 2}}\times \left(R_0^{\alpha  + 2} + \frac{{\alpha r_0^{\alpha  + 2}}}{2}\right)  
   \times \left[f\left(\lambda _{\mathtt{th}}^{\mathtt{off}}\pi R_0^2\right) - f\left(\lambda _{\mathtt{th}}^{\mathtt{off}}\pi \left(R_0^2 - R_s^2\right)\right)\right]  \nonumber \\
& + \lambda _{\mathtt{th}}^{\mathtt{off}} \pi R_s^2d_M^\alpha \times f\left(\lambda _{\mathtt{th}}^{\mathtt{off}}\pi \left(R_0^2 - R_s^2\right)\right),
\end{align}
such that
\begin{itemize}
\item If ${\lambda _0} < \lambda _{\mathtt{th}}^{\mathtt{off}}$, all the SBSs are turned off (i.e., $m_{th}^1=M$);
\item Otherwise, there is at least one SBS should be turned on (i.e., $m_{th}^1<M$).
\end{itemize}
\end{proposition}

\begin{IEEEproof}
 Please refer to Appendix E.
\end{IEEEproof}

\begin{corollary}
The user-density threshold $\lambda _{\mathtt{th}}^{\mathtt{off}}$ decreases with the noise power density $N_0$, the path-loss exponent $\alpha$,  and the required data rate $b$, and increases with the maximum allowable outage probability $\varepsilon$ and the macro-cell operation spectrum bandwidth $W$.
\end{corollary}

\begin{IEEEproof}
 Please refer to Appendix F.
\end{IEEEproof}

\begin{remark}
 Intuitively, as the noise power density $N_0$ and the path-loss exponent $\alpha$ increase, the transmit power of the MBS   increases to compensate the signal loss and it is more power-saving  to use SBSs with a smaller $\lambda_{\mathtt{th}}^{\mathtt{off}}$. The same result holds when the user rate requirement $b$ increases and the maximum allowable outage probability $\varepsilon$ decreases. As $W$ increases, the MBS can use less power to meet the same QoS and thus $\lambda _{\mathtt{th}}^{\mathtt{off}}$ increases.
\end{remark}

The following proposition gives the other macro-cell user density threshold.

\begin{proposition}\label{proposition: 3}
 Under the setup of uniformly distributed users and assuming $P^t_{max}=\infty$, there always exists  a user-density threshold $\lambda _{\mathtt{th}}^{\mathtt{on}}$, which is  the unique solution to
\begin{align}
\label{lamda_th_2_uniform}
 \Delta p&=\left[\frac{{2 \lambda_{\mathtt{th}}^{\mathtt{on}} \pi}}{{\alpha  + 2}}\left(R_0^{\alpha  + 2} + \frac{{\alpha r_0^{\alpha  + 2}}}{2}\right) - \lambda_{\mathtt{th}}^{\mathtt{on}} \pi R_s^2   \sum\limits_{m = 2}^M {d_m^\alpha } \right] 
 \times \left[f\left(\lambda_{\mathtt{th}}^{\mathtt{on}}\pi \left (R_0^2\! - \!(M \!- \!1) R_s^2\right)\right)\! - \! f\left(\lambda_{\mathtt{th}}^{\mathtt{on}}\pi \left(R_0^2 \!- \!M  R_s^2\right)\right)\right]\nonumber   \\
&+ f\left(\lambda_{\mathtt{th}}^{\mathtt{on}}\pi \left(R_0^2 - M  R_s^2\right)\right) \times \lambda_{\mathtt{th}}^{\mathtt{on}} \pi R_s^2d_1^\alpha,
\end{align}
such that
\begin{itemize}
\item If ${\lambda _0} > \lambda _{\mathtt{th}}^{\mathtt{on}}$, all the small-cells should be turned on (i.e., $m_{th}^1=0$);
\item Otherwise, there is at least one small-cell which should be turned off (i.e., $m_{th}^1>0$).
\end{itemize}
\end{proposition}

The proof of Proposition \ref{proposition: 3} is similar to that to Proposition \ref{proposition: 2}, and thus is omitted here for brevity.
In addition, as $N_0$ and $\alpha$ increase, the transmit power of the MBS increases and it is more desirable  to use small-cells with a smaller $\lambda_{\mathtt{th}}^{\mathtt{on}}$. The same result holds when $b$ increases and $\varepsilon$ decreases. As $W$ increases, the MBS's transmit power decreases and $\lambda _{\mathtt{th}}^{\mathtt{on}}$ increases. The relationships between $\lambda _{\mathtt{th}}^{\mathtt{on}}$ and its related parameters are the same as those of the threshold $\lambda _{\mathtt{th}}^{\mathtt{off}}$ in Corollary 1, and thus can be easily proved by using a method similar to the proof   to Corollary~1.

\section{Small-Cell Dynamic Operation for Non-Uniformly Distributed Users}
In this section, we study the more general case where the user densities in the HetNet can be different.  That is, we allow ${\lambda _m} \ne {\lambda _n}$, if $m\neq n$. As compared to the uniform user distribution case in Section III, the HetNet power minimization problem  becomes more complicated in this case. We need to take into account not only the SBSs'  different locations but also the  different user densities in their coverage areas. In the following, we first formulate the HetNet power minimization problem for the case with non-uniformly distributed users, and then present an efficient method to solve this problem sub-optimally in general.

\subsection{Problem Formulation for Non-Uniformly Distributed Users}
Similar to the case with uniformly distributed users, by deciding the  SBSs' optimal operation mode $\bm{\theta}$, we formulate the HetNet power minimization problem   under the case with non-uniformly distributed users as follows: First, the objective is to minimize the  HetNet total power consumption, which is obtained by substituting the MBS's transmit power $P^{t}(\bm{\theta})=T(\bm{\theta})  Z(\bm{\theta})$, given in Theorem 1, into \eqref{p_network}. Second, the problem constraint is given by the MBS's transmit power constraint, i.e.,  $P^{t}(\bm{\theta})=T(\bm{\theta})   Z(\bm{\theta})\leq P_{max}^{t}$.
We refer to  this problem as problem (P2). The form of Problem (P2) is similar to that of  problem (P1)   in Section III,  and thus is omitted here for brevity.  Similar to (P1), we assume $P^{t}(\bm{\theta})=T(\bm{\theta})  Z(\bm{\theta})\leq P_{max}^{t}$ always holds for (P2) to consider a feasible problem.

It is noted that although appearing similarly, problem (P2) is different from  (P1),  since $T({\bm{\theta}} )$ and $Z({\bm{\theta}} )$ in (P2) are given by (\ref{T1}) and (\ref{C1}), respectively. Moreover, in the case with non-uniformly distributed users, besides  the SBSs' different distances to the MBS, the user densities in different SBSs' coverage areas are also different in general.    Thus, as compared to (P1), it is more difficult to solve (P2) optimally.    For example, Proposition 1 for the case with uniformly distributed users does not hold any more for the case with non-uniformly distributed users.  In the following, we first show that problem (P2) is NP-hard in general. Then, we solve (P2)  by jointly considering the impact of SBSs' different distances to the MBS as well as the different user densities in their coverage areas.

\subsection{Computation Complexity of Problem (P2)}
This subsection discusses the computational complexity for solving (P2). In the following, for the ease of analysis, we first rewrite (P2) without loss of optimality. Then we show that (P2) is an NP-hard problem.

We divide the MBS's traffic load into two parts: the first part is the traffic load in $A_0$, given by   $\underline L = {\lambda _0}\pi R_0^2 - {\lambda _0}M\pi R_s^2$, and the second part is the traffic load in all the sleeping SBSs' coverage areas, which is given by
\begin{equation}
\label{L_additional}
L = \sum\limits_{\{ m|{\theta _m} = 0\} } {{\lambda _m}\pi R_s^2}.
\end{equation}
Clearly, $\underline L$ is a constant and $L$ depends on the SBSs' operation modes $\bm{\theta}$.
First, from (\ref{p_network}) and Theorem 1,  it is easy to verify that the total  HetNet   power consumption
by assuming all SBSs are active is given by
\begin{align}
\label{P_inl}
 \widehat P(L) &=  \frac{{u\Gamma {N_0}W}}{{ - D\ln (1 - \varepsilon ) r_0^\alpha}} \times \frac{{{e^{({2^{\frac{b}{W}}} - 1)({\underline L} + L)}} - 1}}{{{\underline L} + L}} \times \Upsilon,
\end{align}
where  $\Upsilon=\frac{{2\pi {\lambda _0}}}{{\alpha  + 2}}(R_0^{\alpha  + 2} + \frac{{\alpha r_0^{\alpha  + 2}}}{2}) - {\lambda _0}\sum\limits_{n = 1}^M {\pi R_s^2d_n^\alpha}$.
Then, if   SBS $m \in \mathcal{M}$ is turned off, the SBSs' total power consumption is  decreased by $\Delta p$, and the MBS's power consumption increases to serve all the users in the  SBS $m$'s coverage area. Denote  the  saved total power consumption in the HetNet by turning off SBS $m$ as ${\Delta _m}(L)$. For any  $m \in \mathcal{M}$,  we have
\begin{align}
\label{P_m_v}
 {\Delta _m}(L)  
=  \Delta p
 - \frac{{u\Gamma {N_0}W}}{{ - D\ln (1 - \varepsilon ) r_0^\alpha}} \times \frac{{{e^{({2^{\frac{b}{W}}} - 1)({\underline L} + L)}} - 1}}{{{\underline L} + L}}{\lambda _m}\pi R_s^2d_m^\alpha.
\end{align}

From (\ref{p_network}), the terms $\underline P$ and $Mp_0$ in problem (P2)'s objective $P^{Het}(\bm{\theta} ) $ are constants, and $M\Delta p$ is also a constant. As a result, it is easy to find that  minimizing $P^{Het}(\bm{\theta} )$ in  problem (P2) is equivalent to minimizing
\begin{equation*}
\label{new_objective_function}
P^{Het}(\bm{\theta} )\!-\!(\underline P\!+\!M{p_0}\!+\!M\Delta p)\!=\!\widehat P (L)\! - \!\!\!\sum\limits_{\{ m|{\theta _m} = 0\} } \!\!{{{\Delta }_m}(L)}.
\end{equation*}
 Therefore, (P2) can be rewritten as
 \begin{align*}
\text{(P2-E)}:\;\mathop {\min }_{\bm{\theta}}  \;\;&\widehat P (L) - \sum\limits_{\{ m|{\theta _m} = 0\} } {{{\Delta }_m}(L)}\nonumber, \\
s.t.\;\;%&\sum\limits_{\{ m|{\theta _m} = 0\} } {{\lambda _m}\pi R_s^2}=L,\nonumber\\
&T({\bm{\theta}} )   Z({\bm{\theta}} ) \le P_{max }^t,\nonumber\\
&{\theta _m} \in \{ 0,\;1\} ,\;m \in \mathcal{M}.
\end{align*}

We now show that (P2-E) and thus (P2) are both NP-hard.  Specifically, we consider a special case of (P2-E) by fixing $L=L_0$, with a given $L_0>0$. In this case, problem (P2-E) becomes the following problem
\begin{align*}
\text{(P2')}:\;\mathop {\min }\limits_{\bm{\theta}}  \;\;&\widehat P (L_0) - \sum\limits_{\{ m|{\theta _m} = 0\} } {{{\Delta }_m}(L_0)}\nonumber, \\
s.t.\;\;&\sum\limits_{\{ m|{\theta _m} = 0\} } {{\lambda _m}\pi R_s^2}=L_0,\nonumber\\
&T({\bm{\theta}} )  Z({\bm{\theta}} ) \le P_{max }^t,\nonumber\\
&{\theta _m} \in \{ 0,\;1\} ,\;m \in \mathcal{M}
\end{align*}
which is an NP-hard Knapsack problem with bag volume $L_0$ and object weights ${{\Delta }_m}(L_0)$'s\cite{Wolsey1999}.
As a result, since (P2-E) is  a general case of (P2') without $L=L_0$, it is easy to   verify that since (P2') is an NP-hard problem, (P2-E) and thus (P2) are also NP-hard problems.

Due to the high complexity to find the optimal solution to (P2), in the following subsections, we first look at some special cases of problem (P2), and then extend the results obtained under the special cases to the general case.

\subsection{ Dynamic Operations of SBSs under Special Cases}
This subsection studies some special cases for problem (P2). In particular, if the traffic load in a small-cell is heavy, it is expected  that this SBS is  active to serve the densely deployed users, regardless of other SBSs' operation modes. Similarly, if a small-cell covers only few users, it is expected that this SBS is inactive to save power. Based on such an observation, we obtain the following proposition.

\begin{proposition}
Two small-cell user density regimes are then given in the following to decide the  optimal operation modes of the SBSs for the case with non-uniformly distributed users.
\begin{itemize}
\item \emph{Low user density regime:} If the small-cell user density $\lambda_m$ in $A_m$ is low with  ${\lambda _m} < \lambda _{\mathtt{th}}^{\mathtt{off}}(m)$, where the threshold $\lambda _{\mathtt{th}}^{\mathtt{off}}(m)$ is the unique solution to
\begin{align}
\label{lamda_th_1_m}
\Delta p&= \left(\Upsilon  + \sum\limits_{n = 1,n \ne m}^M {{\lambda _n}\pi R_s^2d_n^\alpha }  +  \lambda _{\mathtt{th}}^{\mathtt{off}}(m)\pi R_s^2d_m^\alpha \right) \nonumber \\
&\times \Bigg[f\Bigg(\underline L  + \sum\limits_{n = 1,n \ne m}^M\!\!\! {{\lambda _n}\pi R_s^2}
 + \lambda _{\mathtt{th}}^{\mathtt{off}}(m)\pi R_s^2\Bigg)  
 - f\Bigg(\underline L  + \!\!\!\sum\limits_{n = 1,n \ne m}^M \!\!\!{{\lambda _n}\pi R_s^2}\Bigg)\Bigg] \nonumber \\
 & + f\left(\underline L  +\!\!\! \sum\limits_{n = 1,n \ne m}^M {{\lambda _n}\pi R_s^2}\right) \lambda _{\mathtt{th}}^{\mathtt{off}}(m)\pi R_s^2d_m^\alpha,
\end{align}
then it is optimal to turn off  SBS $m$ for problem (P2).
\item \emph{High user density regime:} If the small-cell user density $\lambda_m$ in $A_m$ is high with ${\lambda _m} > \lambda _{\mathtt{th}}^{\mathtt{on}}(m)$, where the threshold $\lambda _{\mathtt{th}}^{\mathtt{on}}(m)$ is the unique solution to
\begin{align}
\label{lamda_th_2_m}
\Delta p =\Upsilon \times \left[f\left(\underline L  + \lambda _{\mathtt{th}}^{\mathtt{on}}(m)\pi R_s^2\right)- f\left(\underline L\right)\right] 
 + f\left(\underline L +  \lambda _{\mathtt{th}}^{\mathtt{on}}(m)\pi R_s^2\right)  \lambda _{\mathtt{th}}^{\mathtt{on}}(m)\pi R_s^2d_m^\alpha,
\end{align}
then it is optimal to turn on SBS $m$.
\end{itemize}
\end{proposition}

 \begin{IEEEproof}
  Please refer to Appendix G.
 \end{IEEEproof}

By using a   method similar  to the proof to Corollary 1, it is also easy to verify that both $\lambda _{\mathtt{th}}^{\mathtt{off}}(m)$ and $\lambda _{\mathtt{th}}^{\mathtt{on}}(m), m \in \mathcal{M}$ decrease with the noise power density $N_0$, the path loss exponent $\alpha$, and the minimum data rate $b$, and increase with the outage probability $\varepsilon$ and macro-cell spectrum bandwidth $W$.

Furthermore, it is easy to find that if the small-cell user density $\lambda_m$ in $A_m$ is extremely low with ${\lambda _m} < \underline \lambda _{\mathtt{th}}^{\mathtt{off}},\forall m \in \mathcal{M}$, where $\underline \lambda _{\mathtt{th}}^{\mathtt{off}}$ is obtained by setting  $m = \arg \mathop {\max }\limits_{n \in \mathcal{M}} {d_n}$, $\lambda _{\mathtt{th}}^{\mathtt{off}}(m)= \underline \lambda _{\mathtt{th}}^{\mathtt{off}}$, and ${\lambda _n} = \underline \lambda _{\mathtt{th}}^{\mathtt{off}}$, $\forall n \in \mathcal{M}$, $n \ne m$ in \eqref{lamda_th_1_m}, it is optimal to turn off all the SBSs.
In this case, all $\lambda_m$'s, $\forall m\in \mathcal{M}$, are within the low user density regime  in Proposition 4. On the other hand, if the small-cell user density $\lambda_m$ in $A_m$ is extremely high with ${\lambda _m} > \overline \lambda _{\mathtt{th}}^{\mathtt{on}}, \forall m \in \mathcal{M}$, where $\overline \lambda _{\mathtt{th}}^{\mathtt{on}}$ is obtained by
setting $m = \arg \mathop {\min }\limits_{n \in \mathcal{M}} {d_n}$,
and $\lambda _{\mathtt{th}}^{\mathtt{on}}(m)= \overline \lambda _{\mathtt{th}}^{\mathtt{on}}$ in \eqref{lamda_th_2_m},
it is optimal to turn on all the SBSs. In this case, all $\lambda_m$'s, $\forall m\in \mathcal{M}$, are within the high user density regime  in Proposition 4.

\subsection{Dynamic Operations of SBSs for General Case}
In this subsection, we extend the location-based operation algorithm proposed for the case with uniformly distributed users for problem (P1) to the case with non-uniformly distributed case and  solve problem (P2). In particular,  by noticing  that in the case with non-uniformly distributed users,   the user density $\lambda_m$ in each small-cell as well as the distance $d_m$ between the SBS $m$ and the MBS are both in general different for different small-cells, and thus are essential for determining the SBSs' operation modes, we propose a location-and-density-based operation algorithm, as given by Algorithm 2, to solve (P2).

\begin{algorithm}[htbp!]
\caption{Location-and-Density-Based Operation Algorithm for the Case with Non-Uniformly Distributed Users}
\label{algorithm_non_uniform}
\begin{small}
\begin{algorithmic}[1]
\STATE $\bm{\theta}_0^* \leftarrow \text{empty set}$, $\bm{\theta}_1^* \leftarrow \text{empty set}$
\FOR {$m \in \mathcal{M}$}
\IF {${\lambda _m} > \lambda _{\mathtt{th}}^{\mathtt{on}}(m)$ given in \eqref{lamda_th_2_m}}
\STATE ${\theta _m} \leftarrow 1$, $\bm{\theta}_1^* \leftarrow \bm{\theta}_1^* \cup \{\theta _m\}$, Eject $m$ from $\mathcal{M}$
\IF {${\lambda _m} < \lambda _{\mathtt{th}}^{\mathtt{off}}(m)$ given in \eqref{lamda_th_1_m}}
\STATE ${\theta _m} \leftarrow 0$, $\bm{\theta}_0^* \leftarrow \bm{\theta}_0^* \cup \{\theta _m\}$, Eject $m$ from $\mathcal{M}$
\ENDIF
\ENDIF
\ENDFOR
\STATE update $M$, $\pi \leftarrow \text{empty set}$
\FORALL{$1 \le m < n \le M$}
\IF{$L_{m=n}$ is feasible}
\STATE $\pi  \leftarrow \pi  \cup \{ L_{m=n}\}$
\ENDIF
\ENDFOR
\STATE $\pi  \leftarrow \pi  \cup \{ 0,\;\sum\limits_{m = 1}^{M} {{\lambda _m}\pi R_s^2} \}$
\STATE $N \leftarrow |\pi |$
\STATE Reorder $\pi$ in ascending order
\STATE Reorder small-cells as ${Q_m}(0)$'s decrease
\STATE ${\bm{\theta}_a} = [1,1, \cdots ,1]$
\FOR{$n = 2:N$}
\FORALL{$l \in [1,\cdots,M]$ s.t. $\pi (n - 1) \le \sum\limits_{m = 1}^l {{\lambda _m}\pi R_s^2}  \le \pi (n)$}
\STATE ${\bm{\theta}_t} = [\bm{0}_{1 \times (l-1)},{\theta _l} = 0,\bm{1}_{1 \times (M-l)}]$
\IF {$P^{Het}([\bm{\theta}_0^*,\bm{\theta}_t,\bm{\theta}_1^*])<P^{Het}([\bm{\theta}_0^*,\bm{\theta}_a,\bm{\theta}_1^*])$ and $P^t([\bm{\theta}_0^*,\bm{\theta}_t,\bm{\theta}_1^*]) \le P_{max}^t$}
\STATE ${\bm{\theta}_a} \leftarrow \bm{\theta}_t$
\ENDIF
\ENDFOR
\STATE Perform the switching corresponding to $\pi (n)$
\ENDFOR
\STATE RETURN $[\bm{\theta}_0^*,\bm{\theta}_a,\bm{\theta}_1^*]$
\end{algorithmic}
\end{small}
\end{algorithm}

Specifically, to solve problem (P2) in the general case, similar to Algorithm 1 proposed  for solving (P1), by assume all SBSs are initially active, we select some SBSs and gradually  turn off them. However, unlike the uniform user distribution case in Section III,  it is difficult to find a proper order for all the SBSs so as to turn off the SBSs by the distance order to the MBS   in the non-uniform user distribution case,  since such a SBS order depends on  the traffic load $L$ in the inactive SBSs to be selected, given in (\ref{L_additional}). It is also noted that  $L$ is determined by the SBSs' operation modes $\bm{\theta}$. As a result, from  (\ref{L_additional}), we can find that depending on the SBSs' operations modes,  $L$ can be any value in the range that is given by $\left[0,\sum\limits_{m = 1}^M {{\lambda _m}\pi R_s^2}\right]$. We refer to such a range  as the \emph{feasible range of $L$} in the following. For any given $L$ in its feasible range,  define ${Q_m}(L) = \frac{{\Delta_m(L)}}{{{\lambda _m}\pi R_s^2}}
$ as the \emph{power-saving efficiency} if SBS $m \in\mathcal{M}$ is turned off, which gives the saved power consumption by turning off SBS $m$ per  unit load. To save the HetNet total power consumption, it is preferred to turn off SBSs with large ${Q_m}(L)$.

Our solution approach in Algorithm 2 for problem (P2) is now given as follows: i) Given any $L$, we find all the power-saving efficiencies ${Q_m}(L), m \in \mathcal{M}$ and reorder SBSs as their $Q_m(L)$'s decrease. We name such a SBS order as the ``power-saving list''; ii) Since the ``power-saving list'' changes with $L$, we search over $L\in \left[0,\sum\limits_{m = 1}^M {{\lambda _m}\pi R_s^2}\right]$ to find all the ``power-saving lists''. For every ``power-saving list'', we also find its feasible range of $L$; iii) According to every ``power-saving list'', starting with all small-cells active, we gradually deactivate small-cells. In every deactivation step, we check the traffic that is handed over to the macro-cell. If it is within the above mentioned feasible range of $L$, we record the SBSs' operation modes in this deactivation step as a ``candidate solution''. Our scheme is to choose the best solution among all the ``candidate solutions''.

The key of the above method is   to find all the ``power saving lists'' and the corresponding feasible ranges of $L$, which can be obtained by using an efficient method as shown below. Given any two SBSs $m$ and $n$, the relationship between ${Q_m}(L)$ and ${Q_n}(L)$  can be either of  the following two cases. One   is that ${Q_m}(L)$ (or ${Q_n}(L)$) is always larger than ${Q_n}(L)$ (or ${Q_m}(L)$) over the whole range of $L\in \left[0,\sum\limits_{m = 1}^M {{\lambda _m}\pi R_s^2}\right]$; the other   is that ${Q_m}(L)>{Q_n}(L)$ (or ${Q_n}(L)>{Q_m}(L)$) becomes ${Q_n}(L)>{Q_m}(L)$ (or ${Q_m}(L)>{Q_n}(L)$) after some ``switching point'' at $L\in \left[0,\sum\limits_{m = 1}^M {{\lambda _m}\pi R_s^2}\right]$, which is denoted as $L_{m=n}$. For the latter case, the ``switching point'' $L_{m=n}$ of any pair of SBSs $m$ and $n$ is the unique solution to
${Q_m}(L_{m=n}) = {Q_n}(L_{m=n})$
with  $L_{m=n}\in \left[0,\sum\limits_{m = 1}^M {{\lambda _m}\pi R_s^2}\right]$.
As a result,  for any  two SBSs $m$ and $n$, we can first find the ``switching point'' $L_{m=n}$, and after finding all the ``switching points'' for every two SBSs, we reorder these ``switching points'' according to their values in ascending order. It can be easily verified  that there exists at most one ``switching point''  for any two SBSs; thus, the number of ``switching points'' is $\mathcal{O}(M^2)$. Next, we reorder the SBSs according to  ${Q_m}(0)$'s decreasing order to obtain the initial ``power-saving list''. Starting from the minimum ``switching point'', for every ``switching point'' $L_{m=n}$, we switch the orders of small-cells $m$, $n$ to obtain a new ``power-saving list''. The number of ``power-saving lists'' is also $\mathcal{O}(M^2)$.
It is easy to find that according to each ``power-saving list'',  the computation order  of the above mentioned searching process is $\mathcal{O}(M)$. The computation order of  Algorithm \ref{algorithm_non_uniform} is thus $\mathcal{O}(M^3)$.

\section{Simulation Results}
This section presents the simulation results to study the performance of the proposed HetNet power saving scheme. We focus on the general case with non-uniformly distributed users and investigate Algorithm 2. We also observe similar performance for Algorithm 1 for the special case with uniformly distributed users, which is thus omitted here for brevity. For the MBS's total power consumption, which is given in  \eqref{P_macro}, we set $\underline P=712$W, $u=14.5$, and $P_{max}^t=40$W \cite{D2.3}. For the small-cells,  we let the  user densities  ${\lambda _m}$'s, $\forall m \in \mathcal{M}$, be uniformly and independently distributed   within the range $[50{\lambda _0} - \sqrt {3{\sigma ^2}} ,50{\lambda _0} + \sqrt {3{\sigma ^2}}]$, where ${\sigma ^2}$ is the user-density variation.  We also follow \cite{Che.JSAC.15} to consider a practically large range for $\lambda_0$.
 Other simulation parameters are set as follows: the capacity loss $\Gamma  = 1$, the fixed path loss $D=-35$dB, the reference distance $r_0=1$m, the path-loss component $\alpha=2.5$,  the bandwidth $W=10$ MHz,   the required rate $b=0.1$Mbits/sec, the maximum allowable outage probability ${\varepsilon}=0.05$, and the  noise power ${N_0} =  - 174$dBm/Hz.
 In the following, we first show the near-optimal performance of  the proposed Algorithm \ref{algorithm_non_uniform}. Then we compare our proposed HetNet power saving scheme with two benchmark schemes, where one is without SBS on/off adaptation, and the other is  probability-based SBS on/off adaptation. At last,  we show the individual power consumption of the MBS and all the SBSs, to  further understand the HetNet  power saving problem.

\subsection{Near-Optimal Performance of Algorithm 2}
In this subsection, we compare the  performance of Algorithm \ref{algorithm_non_uniform} with that of  the optimal solution to problem (P2), which  is obtained by  exhaustive search.  It is easy to verify that the computational complexity of the exhaustive search is  $\mathcal{O}(2^M)$. Due to such an   exponentially increased computational complexity over $M$, we only consider a scenario with $M=20$.  As shown in Table II, we calculate the ratio of the HetNet's total power consumption that is obtained by the optimal  exhaustive search over that  obtained by Algorithm \ref{algorithm_non_uniform}, i.e.,  $\frac{{P^{Het-opt}}}{{P^{Het-Alg2}}}$. It is observed  that the performance of the proposed Algorithm \ref{algorithm_non_uniform} achieves less than $1\%$ performance loss as compared to the optimal one over all macro-cell user density $\lambda_0$ in Table II.

\begin{table}
\tabcolsep 1.5mm \caption{Performance Comparison of Algorithm 2 with the Optimal Solution to Problem (P2)}
\begin{center}
\begin{tabular}{|c|c|c|c|c|c|c|c|c|}
  \hline
  ${\lambda _0}~(\times 10^{-3}) /\textrm{m}^2$&$0.2$&$0.4$&$0.6$&$0.8$&$1.0$&$1.2$&$1.4$&$1.6$\\
  \hline
  \hline
  $\frac{{P^{Het-opt}}}{{P^{Het-Alg2}}}$&$0.9996$&$0.9986$&$0.9978$&$0.9973$&$0.9978$&$0.9982$&$0.9988$&$0.9995$\\
  \hline
\end{tabular}
\end{center}%\vspace{-30pt}
\end{table}

\subsection{Comparison with Benchmark Schemes}
\begin{figure}
\centering
\DeclareGraphicsExtensions{.eps,.mps,.pdf,.jpg,.png}
\DeclareGraphicsRule{*}{eps}{*}{}
\includegraphics[angle=0, width=0.7\textwidth]{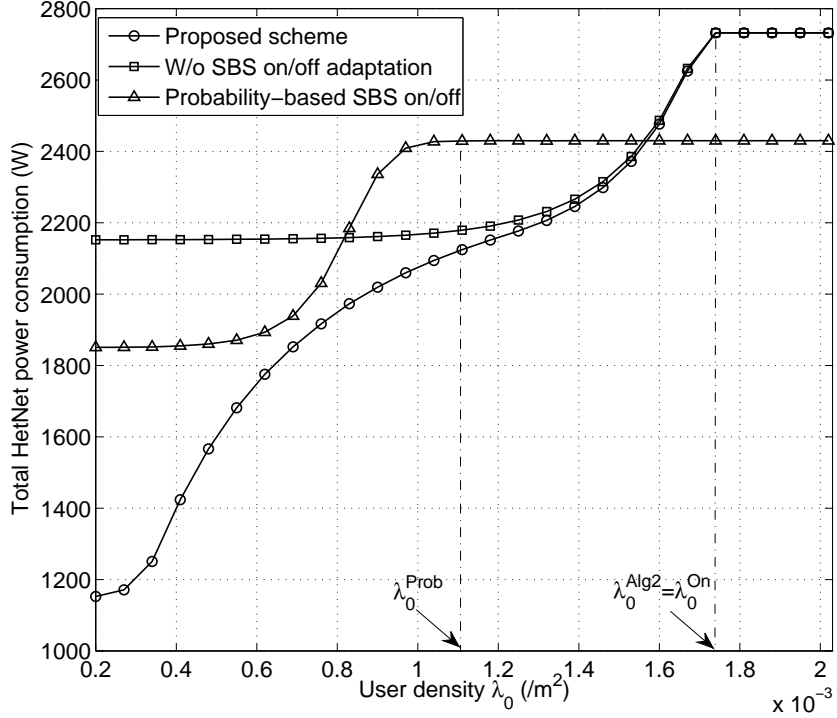}
\caption{Comparison with the benchmark schemes. }
\label{fig: compare}
%\vspace{-0.1in}
\end{figure}

In this subsection, to further show  the performance of the proposed HetNet power saving scheme, we compare  Algorithm~\ref{algorithm_non_uniform} with the following two benchmark schemes. One is without  on/off operation   adaptation for all SBSs,  and keep all SBSs always active. The other is to independently keep all SBSs active based on a probability $P_{active}$. We set $P_{active}=0.7$ in the simulation. For all three schemes, we consider  a case with dense small-cell deployment with $M=144$. In Fig.~\ref{fig: compare}, we show the  total HetNet  power consumption across all SBSs and the MBS, given in (\ref{p_network}),  for  all three schemes over a large range of macro-cell user density $\lambda_0$. For each of the three schemes,    the MBS adopts the same power allocation method as that discussed in Section II-C to assure the macro-cell users' QoS.
As has been discussed in Section III-A, if $\lambda_0$ is too large, such that the MBS's maximum transmit power $P_{max}^{t}$ cannot satisfy all macro-cell users' QoS, the system becomes infeasible. In this case, for all three schemes, we apply admission control to  randomly reject a portion of users  from accessing the HetNet such that the MBS's
maximum transmit power is just sufficient to assure the QoS of all the  remaining users that are allowed to access the HetNet.   We denote the macro-cell user density that makes the resulted MBS transmit power  equal to $P_{max}^{t}$ for our proposed scheme, the scheme without SBS on/off, and the scheme with probability-based SBS on/off as $\lambda_0^{Alg2}$, $\lambda_0^{On}$, and $\lambda_0^{Prob}$, respectively.

It is first observed from Fig.~\ref{fig: compare} that the HetNet total  power consumptions of all three schemes are non-decreasing over $\lambda_0$ as expected.
It is also observed that our proposed scheme always consumes much less power than the  other two benchmark schemes when $\lambda_0$ is  not large, and gradually approaches to the scheme without SBS on/off  as $\lambda_0$ increases, where  increasingly more active SBSs are needed under our proposed scheme to support the increased users. When all SBSs under our proposed scheme become active, our proposed scheme becomes the scheme without SBS on/off, and thus consumes the same power consumption.
Moreover, when $\lambda_0>\lambda_0^{Alg2}=\lambda_0^{On}$, the system becomes infeasible for both proposed scheme and the  scheme without SBS on/off. In this case,
since the MBS keeps using the maximum transmit power $P_{max}^{t}$ and the SBSs are always active,  the HetNet total  power consumption under both schemes becomes a constant, which is given by $\underline{P} + uP^t_{max}+144p_0+144  \times (p_1-p_0)=2732$W from (\ref{p_network}).  Similarly, when  $\lambda_0> \lambda_0^{Prob}$, the system also becomes infeasible under the probability-based scheme, where the HetNet  total  power consumption is given by $\underline{P} + uP^t_{max}+144p_0+144\times P_{active} \times (p_1-p_0)=2429.6$W, which is smaller than that under the proposed scheme since only $100P_{active}\%$ of SBSs are active. Furthermore, since $\lambda_0^{Alg2}> \lambda_0^{Prob}$, our proposed scheme can properly support more users than the  probability-based scheme.

\subsection{Power Consumption of MBS and All SBSs}

\begin{figure}
\centering
\DeclareGraphicsExtensions{.eps,.mps,.pdf,.jpg,.png}
\DeclareGraphicsRule{*}{eps}{*}{}
\includegraphics[angle=0, width=0.7\textwidth]{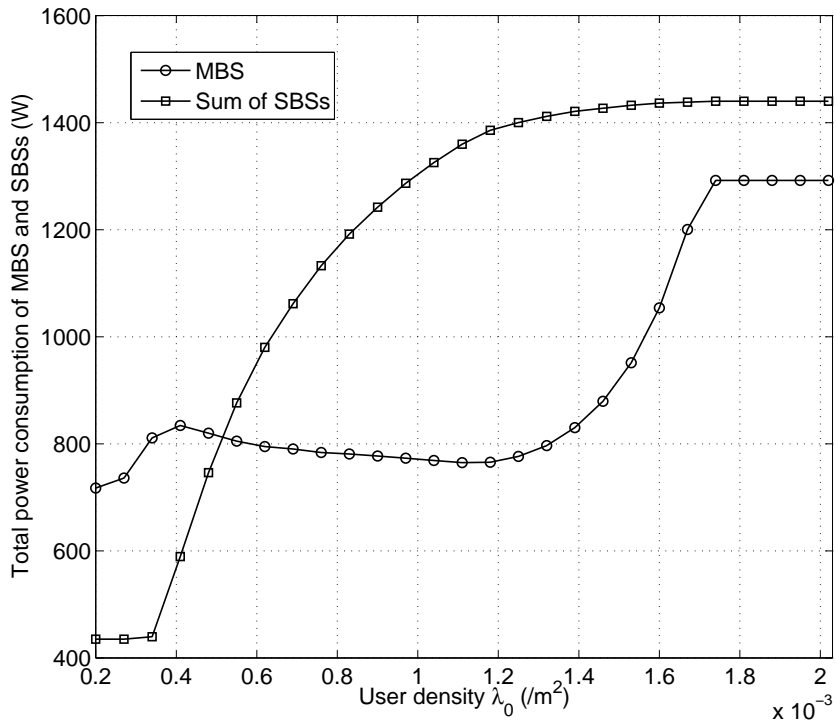}
\caption{ Total power consumption of the MBS and the SBSs.  }
\label{fig: power}
%\vspace{-0.1in}
\end{figure}

In this subsection, we show the  total power consumption of the MBS as well as that of all the SBSs  by applying Algorithm 2. The simulation parameters are set as the same as those in Fig.~\ref{fig: compare}.  As shown in Fig.~\ref{fig: power}, it is first observed that when $\lambda_0$ is small, since all the users in the HetNet can be properly supported by the MBS, the SBSs just remain inactive with a constant power consumption given by $144\times p_0=432$W, while the MBS's power consumption increases over $\lambda_0$. As $\lambda_0$ increases, the SBSs begin to be turned on and thus their total  power consumption begins to increase over  $\lambda_0$. It is interesting to observe from  Fig.~\ref{fig: power} that when $\lambda_0$ is between $[0.4,1.2]\times 10^{-3}$, the MBS's power consumption becomes decreasing over $\lambda_0$, due to the efficient traffic offloading from the SBSs that decreases the MBS's power consumption. However, when $\lambda_0$ is sufficiently large with $\lambda_0>1.2\times 10^{-3}$, such that  the SBSs'
power consumption for offloading MBS's traffic load can cause the overall HetNet power to  increase, we observe that the MBS needs to increase its transmit power again to properly support the users for  saving the total power consumption in the HetNet. At last, similar to that in Fig.~\ref{fig: compare}, when the traffic load in the HetNet is unexpectedly heavy, all the SBSs become active and the MBS transmits with its maximum power. In this case, we  observe that the total power consumption of the MBS is  a constant, given by $\underline{P} + uP^t_{max}=1292$W, and that of all the SBSs is also a constant, given by  $144\times p_1=1440$W.

\section{Conclusion}
In this paper, by dynamically changing SBSs'  operation  modes, we studied the HetNet power minimization problem for both cases with uniformly and non-uniformly distributed users.   For the case with uniformly distributed users, the SBS's location-based operation algorithm was proposed and the optimal operation mode can be found by gradually deactivating SBSs according to their individual distances to the MBS. For the case with non-uniformly distributed users, both SBSs' different locations and small-cell user density were considered for their optimal operation mode decision. We showed that the optimal operation in this case is NP-hard and then proposed a near-optimal algorithm with polynomial complexity. In the future work, it is interesting to extend the current single-macro-cell scenario to the general multi-macro-cell scenario by further taking the interference from neighboring  macro-cells into account.

\appendices

\section{Proof of Theorem 1}
We first derive the inner expectation of  $\mathtt{E}_{r_k}\left[P_k^t({r_k},K)\right]$ of (\ref{P_t_avg}), for a given $K$.
Let $1_{\{k\in A_n\}}$, $\forall n\in \{0\}\cup\mathcal{M}$, denote whether the macro-cell user $k$ is located in the area $A_n$,  where $1_{\{Y\}}=1$ if the event $Y$ is true, and $1_{\{Y\}}=0$ otherwise. Since each user $k$ is independently distributed in the HetNet and uniformly distributed within each $A_n$, it is easy to verify that $\mathtt{P}\left(1_{\{k\in A_n\}}=1\right)= \frac{\lambda_n \|A_n\|}{\sum_{i=0}^{M}\lambda_i\|{A_i}\| } $. Moreover, given that   user $k$ is located in $A_n$, we can easily obtain
$\mathtt{E}_{r_k}\left[ P_k^t({r_k},K) \Big| 1_{\{k \in A_n\}}=1\right]=\int_{A_n} \frac{1}{\|A_n\|} {P_k^t(r,K)dS} $ since the user is uniformly distributed in $A_n$.
As a result, by considering all the possible events with user $k$ locating in $A_n$, $n=0, \textrm{or~}n\in \mathcal{M},\theta_n=0$, we can write   the inner expectation $\mathtt{E}_{r_k}\left[P_k^t({r_k},K)\right]$ of (\ref{P_t_avg}) as
 \begin{align}
 \mathtt{E}_{r_k}\left[P_k^t({r_k},K)\right]&=\sum_{n=0, \textrm{or~}n\in \mathcal{M},\theta_n=0}\mathtt{P}\left(1_{\{k\in A_n\}}=1\right)\mathtt{E}_{r_k}\left[ P_k^t({r_k},K) \Big| 1_{\{k \in A_n\}}=1\right]\nonumber \\
 &=  \sum_{n=0, \textrm{or~}n\in \mathcal{M},\theta_n=0} \frac{\lambda_n \|A_n\|}{\lambda_0\|{A_0}\| + \sum_{\{ m|{\theta _m} = 0\}} \lambda_m \|{A_m}\|}  \int_{A_n} \frac{1}{\|A_n\|} {P_k^t(r,K)dS}  \nonumber \\
 &=\frac{1}{{{\lambda _0}||{A_0}|| + \sum\limits_{\{ m|{\theta _m} = 0\} } {{\lambda _m}||{A_m}||} }} \cdot \left({{\lambda _0}\int\limits_{{A_0}} {P_k^t(r_k,K)dS}  + \sum\limits_{\{ m|{\theta _m} = 0\} } {{\lambda _m}\int\limits_{{A_m}} {P_k^t(r_k,K)dS} } }\right). \label{eq: inner_expect}
 \end{align}

Next, by calculating  the outer expectation in \eqref{P_t_avg}, we derive $P^t(\bm{\theta})$. Let
\begin{align*}
%\label{g_r}
g(r) = \left\{ {\begin{array}{*{20}{l}}
{\frac{r^\alpha}{r_0^\alpha},} &{ \text{if}\;{r} \ge {r_0},}\\
{1,} & {\text{ otherwise,}}
\end{array}} \right.
\end{align*}
By substituting \eqref{P_n_t} and \eqref{eq: inner_expect} into \eqref{P_t_avg} and noting that $K$ follows Poisson distribution with mean $\mu=\lambda_0\|A_0\|+\sum_{\{m|\theta_m=0\}}\lambda_m \|A_m\|$, it is easy to verify that  ${P^t}({\bm{\theta }}) =Z({\bm{\theta }}) \cdot T({\bm{\theta }})$  with
\begin{align*}
T({\bm{\theta }})= \frac{{\Gamma {N_0}W}}{{ - \ln (1 - \varepsilon )D}}\bigg(\exp\big(({2^{\frac{b}{W}}} - 1)({\lambda _0}\pi R_0^2- {\lambda _0}M\pi R_s^2 + \sum\limits_{\{ m|{\theta _m} = 0\} } {{\lambda _m}\pi R_s^2} )\big) - 1\bigg),
\end{align*}
\begin{align}
Z({\bm{\theta }}) = &\frac{\frac{{2\pi {\lambda _0}}}{{(\alpha  + 2)r_0^{\alpha}}}(R_0^{\alpha  + 2} + \frac{{\alpha r_0^{\alpha  + 2}}}{2}) -\sum\limits_{\{ m|{\theta _m} = 0\} } {{\lambda _0}\int\limits_{{A_m}} {g(r)dS} }+ \sum\limits_{\{ m|{\theta _m} = 0\} } {{\lambda _m}\int\limits_{{A_m}} {g(r)dS} } }{{{\lambda _0}\pi R_0^2 - {\lambda _0}M\pi R_s^2 + \sum\limits_{\{ m|{\theta _m} = 0\} } {{\lambda _m}\pi R_s^2} }}\nonumber\\
\overset{(a)}{=} &\frac{\frac{{2\pi {\lambda _0}}}{{\alpha  + 2}}(R_0^{\alpha  + 2} + \frac{{\alpha r_0^{\alpha  + 2}}}{2}) - {\lambda _0}\sum\limits_{m = 1}^M {\pi R_s^2d_m^\alpha }+\sum\limits_{\{ m|{\theta _m} = 0\} } {{\lambda _m}\pi R_s^2d_m^\alpha }}{{r_0^\alpha ({\lambda _0}\pi R_0^2 - {\lambda _0}M\pi R_s^2 + \sum\limits_{\{ m|{\theta _m} = 0\} } {{\lambda _m}\pi R_s^2)} }}, \label{eq: Z_theta}
\end{align}
where the equality $(a)$ follows due to  the fact that the macro-cell's radius $R_0$ is much larger than the small-cells' $R_s$ and the reference distance $r_0$ in practical systems. Theorem 1 thus follows.

\section{Proof of Proposition 1}
We prove this proposition by the method of reduction to absurdity.
 We assume that in the optimal operation mode $\bm{\theta}^*=[\theta_1^*, \cdots,\theta_M^*]$, a SBS $m \in \mathcal{M}$ is turned off ($\theta_m^*=0$) and a SBS  $n \in \mathcal{M}$ is turned on, with $d_n<d_m$. We now construct a   new solution   as ${\hat{\bm{\theta}} ^{*}} = [\hat{\theta}_1^{*} = \theta _1^{*}, \cdots ,\hat{\theta}_n^{*} = 0, \cdots ,\hat{\theta}_m^{*} = 1, \cdots ,\hat{\theta} _M^{*} = \theta _M^{*}]$. It can be easily verified that $T({\hat{\bm{\theta}} ^{*}})=T({\bm{\theta} ^{*}})$ and $Z({\hat{\bm{\theta}} ^{*}})<Z({\bm{\theta} ^{*}})$, then $P^{Het}({\hat{\bm{\theta}} ^{*}})<P^{Het}({\bm{\theta} ^{*}})$, which conflicts with the optimality of ${\bm{\theta} ^{*}}$. Proposition 1  thus follows.

\section{Proof of Lemma 1}
We first rewrite the MBS's transmit power as $P^t(\bm{\theta})=P^t([\bm{0}_{1 \times (m-1)},{\theta _m} = 0,\bm{1}_{1 \times (M-m)}])= {h_1}(m){h_2}(m)$, where
\begin{align*}
{h_1}(m) =\frac{{\Gamma {N_0}W}}{{ - D\ln (1 - \varepsilon ) r_0^\alpha}} \cdot \frac{{\exp \big(({2^{\frac{b}{W}}} - 1)\lambda {}_0\pi (R_0^2 - MR_s^2 + mR_s^2)\big) - 1}}{{\pi (R_0^2 - MR_s^2 + mR_s^2)}}
\end{align*}
and
\begin{align*}
{h_2}(m) = \frac{{2\pi }}{{\alpha  + 2}}(R_0^{\alpha  + 2} + \frac{{\alpha r_0^{\alpha  + 2}}}{2}) - \pi R_s^2 \cdot \sum\limits_{n = m + 1}^M {d_n^\alpha }.
\end{align*}

Next, to prove this lemma, we verify the following two facts:
\begin{itemize}
\item Fact \#1: ${h_1}(m) - {h_1}(m - 1) > {h_1}(m - 1) - {h_1}(m - 2) > 0$;
\item Fact \#2: ${h_2}(m) - {h_2}(m - 1) \ge {h_2}(m - 1) - {h_2}(m - 2) > 0$.
\end{itemize}

To prove Fact \#1, we suppose that ${h_1}(m)$ is a continuous function of $m$. It can be verified that both of its first-order and  second-order derivatives are positive when $m \ge 0$. Then, the function $h_1(m)$ is an increasing convex function and thus we can easily obtain Fact \#1.
To prove Fact \#2, it can be calculated that ${h_2}(m) - {h_2}(m - 1) = \pi R_s^2d_m^\alpha$ and ${h_2}(m - 1) - {h_2}(m - 2) = \pi R_s^2d_{m - 1}^\alpha$. Noting that ${d_m} \ge {d_{m - 1}}$, we can thus obtain Fact \#2.

Now, we  prove Lemma 1. From Fact \#2 we can obtain that
\begin{align}
\label{un_eqa_1}
&{h_1}(m)\big({h_2}(m) - {h_2}(m - 1)\big) > {h_1}(m - 1)\big({h_2}(m - 1) - {h_2}(m - 2)\big) \Leftrightarrow \nonumber\\
&{h_1}(m){h_2}(m) - {h_1}(m - 1){h_2}(m - 1) > {h_1}(m){h_2}(m - 1) - {h_1}(m - 1){h_2}(m - 2).
\end{align}
Similarly, from  Fact \#1 we can have
\begin{align}
\label{un_eqa_2}
&{h_2}(m - 1){h_1}(m) - {h_2}(m - 2){h_1}(m - 1) > {h_2}(m - 1){h_1}(m - 1) - {h_2}(m - 2){h_1}(m - 2).
\end{align}
Combining  \eqref{un_eqa_1}, and \eqref{un_eqa_2} , we can obtain that
\begin{align*}
%\label{delta_p_increase_temp}
&P^t([\bm{0}_{1 \times (m-1)},{\theta _m} = 0,\bm{1}_{1 \times (M-m)}]) - P^t([\bm{0}_{1 \times (m-2)},{\theta _{m-1}} = 0,\bm{1}_{1 \times (M-m+1)}]) > \nonumber\\
&P^t([\bm{0}_{1 \times (m-2)},{\theta _{m-1}} = 0,\bm{1}_{1 \times (M-m+1)}]) - P^t([\bm{0}_{1 \times (m-3)},{\theta _{m - 2}} = 0,\bm{1}_{1 \times (M-m+2)}]),
\end{align*}
i.e., $\Delta P_m>\Delta P_{m-1}$. Lemma 1 thus follows.

\section{Proof of Theorem 2}
We consider two cases as follows: 1) Case 1: $m_{th}^1\leq m_{th}^2$, and 2) Case 2: $m_{th}^1> m_{th}^2$.
We first look at Case 1.  According to Proposition 1, starting with all SBSs being active, it is optimal to turn off closer SBSs according to their individual distances to the MBS. Suppose this deactivation process stops at SBS $m_{th}$ with $m_{th}<m_{th}^1$. Next,  if we continue to turn off SBS $m_{th}+1$, according to (\ref{delta_p_m}) and Lemma 1, the saved HetNet power is
$\Delta p - \Delta {P_{{m_{th}}+1}} \ge \Delta p- \Delta {P_{m_{th}^1}}>0,
$
which contradicts with optimal stopping at $m_{th}$. With the same method, it is easy to verify that  stopping at any $m_{th}>m_{th}^1$  cannot be optimal. Thus, it is optimal to choose $m_{th}=m_{th}^1$. Note that the transmit power of the MBS is always no larger than $P_{max}^t$ thanks to $m_{th}^1\leq m_{th}^2$.

We now consider Case 2. According to (\ref{m_th_2}), to let the transmit power constraint hold, the optimal stopping SBS's index   $m_{th}$ is  no greater than $m_{th}^2$. If $m_{th}<m_{th}^2$, continuing to turn off small-cell $m_{th}+1$ helps save the HetNet's power by
$
\Delta p - \Delta{P_{{m_{th}}+1}} \ge \Delta p - \Delta {P_{m_{th}^2}} > \Delta p- \Delta {P_{m_{th}^1}}>0.
$
Thus, the only optimal $m_{th}$ is equal to $m_{th}^2$.
As a result, based on the above two cases, it is optimal to deactivate closer small-cells until reaching small-cell $m_{th}=\min(m_{th}^1, m_{th}^2)$. Theorem 2 thus follows.

\section{Proof of Proposition 2}
First, we  prove  the existence and uniqueness of $\lambda _{\mathtt{th}}^{\mathtt{off}}$. Denote the left-hand side of \eqref{lamda_th_1_uniform} as $h_3(\lambda _{\mathtt{th}}^{\mathtt{off}})$. It is easy to verify that $h_3(\lambda _{\mathtt{th}}^{\mathtt{off}})$ is an increasing function over $\lambda _{\mathtt{th}}^{\mathtt{off}} \ge 0$. Since $\mathop {\lim }\limits_{\lambda _{\mathtt{th}}^{\mathtt{off}} \to  + \infty } h_3(\lambda _{\mathtt{th}}^{\mathtt{off}}) \to  + \infty$ and $h_3(0) = 0$, $\lambda _{\mathtt{th}}^{\mathtt{off}}$ exists and it is unique. The solution $\lambda _{\mathtt{th}}^{\mathtt{off}}$ can thus be obtained through bisection search.

Now, based on the unique  $\lambda _{\mathtt{th}}^{\mathtt{off}}$, we prove Proposition 2.
For the case with ${\lambda _0} < \lambda _{\mathtt{th}}^{\mathtt{off}}$, we have
\begin{small}
\begin{align}
\label{uniform_th_low_2}
&P^{Het}([\bm{0}_{1 \times M}]) - P^{Het}([\bm{0}_{1 \times (m-1)},{\theta _m} = 0,\bm{1}_{1 \times (M-m)}])\nonumber\\
\overset{(a)}{=}& \frac{{2 \lambda_0 \pi }}{{\alpha  + 2}}(R_0^{\alpha  + 2} + \frac{{\alpha r_0^{\alpha  + 2}}}{2}) \cdot \bigg(f\big(\lambda_0\pi R_0^2\big) - f\big(\lambda_0\pi (R_0^2 - MR_s^2 + mR_s^2)\big)\bigg)+  f\big(\lambda_0\pi (R_0^2 - MR_s^2 + mR_s^2)\big) \cdot \nonumber\\& \lambda_0 \pi R_s^2 \sum\limits_{n = m + 1}^M {d_n^\alpha }  - (M - m)\Delta p\nonumber\\
 \overset{(b)}{\le} & (M - m)\frac{{2 \lambda _0 \pi }}{{\alpha  + 2}}(R_0^{\alpha  + 2} + \frac{{\alpha r_0^{\alpha  + 2}}}{2}) \cdot \bigg(f\big({\lambda _0}\pi R_0^2\big) - f\big({\lambda _0}\pi (R_0^2 - R_s^2)\big)\bigg)+ (M - m)f\big({\lambda _0}\pi (R_0^2 - R_s^2)\big) \cdot \nonumber\\& \lambda _0 \pi R_s^2d_M^\alpha  - (M - m)\Delta p < 0
\end{align}
\end{small}where $(a)$ follows since $f(x)$ in \eqref{C_fun}  is increasing over  $x\in (0, + \infty )$, and $(b)$ follows by noticing  ${\lambda _0} < \lambda _{\mathtt{th}}^{\mathtt{off}}$, the increasing  of $h_3(\lambda_0)$ over $\lambda_0$, and \eqref{lamda_th_1_uniform}.
Thus, we have $P^{Het}([\bm{0}_{1 \times M}]) - P^{Het}([\bm{0}_{1 \times (m-1)},{\theta _m} = 0,\bm{1}_{1 \times (M-m)}])<0$ for any ${\lambda _0} < \lambda _{\mathtt{th}}^{\mathtt{off}}$ and $0 \le m < M$.
For the case with ${\lambda _0} > \lambda _{\mathtt{th}}^{\mathtt{off}}$. Since $h_3(\lambda_0)$ is increasing over $\lambda_0$, from \eqref{lamda_th_1_uniform}, we can obtain that
 if ${\lambda _0} > \lambda _{\mathtt{th}}^{\mathtt{off}}$, $P^{Het}([\bm{0}_{1 \times M}]) > P^{Het}([\bm{0}_{1 \times (M-1)},1])$. This shows that it is not the optimal solution to close all the SBSs. As a result,   at least one SBS should be active.  Proposition 2 thus follows.

\section{Proof of Corollary 1}
We denote the left-hand side of \eqref{lamda_th_1_uniform} as ${h_3}({N_0},\alpha ,b,\varepsilon ,W,\lambda _{\mathtt{th}}^{\mathtt{off}})$, which  is   a function of $N_0$, $\alpha$, $b$, $\varepsilon$, $W$, $\lambda _{\mathtt{th}}^{\mathtt{off}}$. We use this function to establish the relationships of $\lambda _{\mathtt{th}}^{\mathtt{off}}$ with $N_0$, $\alpha$, $b$, $\varepsilon$, and $W$.
It can be shown that ${h_3}({N_0},\alpha ,b,\varepsilon ,W,\lambda _{\mathtt{th}}^{\mathtt{off}})$ increases with $\lambda _{\mathtt{th}}^{\mathtt{off}}$. It is easy to verify that if  ${h_3}({N_0},\alpha ,b,\varepsilon ,W,\lambda _{\mathtt{th}}^{\mathtt{off}})$ also increases over a parameter   among $N_0$, $\alpha$, $b$, $\varepsilon$, and $W$,    $\lambda _{\mathtt{th}}^{\mathtt{off}}$ decreases with this parameter; otherwise, the opposite is true. For example, it can be easily shown that ${h_3}({N_0},\alpha ,b,\varepsilon ,W,\lambda _{\mathtt{th}}^{\mathtt{off}})$ increases as $b$, $N_0$ increase and $\varepsilon$ decreases. Thus, $\lambda _{\mathtt{th}}^{\mathtt{off}}$ decreases with $b$ and $N_0$, and increases with $\varepsilon$.

We now study how  $\lambda _{\mathtt{th}}^{\mathtt{off}}$ varies over  $\alpha$ and $W$. The partial derivative of ${h_3}({N_0},\alpha ,b,\varepsilon ,W,\lambda _{\mathtt{th}}^{\mathtt{off}})$ with $\alpha$ can be easily shown to be positive, using the fact that the macro-cell's radius $R_0$ and the distance from the MBS to its most far-away small-cell $d_M$ is typically much larger than the reference distance $r_0$. Thus, ${h_3}({N_0},\alpha ,b,\varepsilon ,W,\lambda _{\mathtt{th}}^{\mathtt{off}})$ increases with $\alpha$, then $\lambda _{\mathtt{th}}^{\mathtt{off}}$ decreases with it. To calculate the partial derivative with $W$, we fix other variables and rewrite ${h_3}({N_0},\alpha ,b,\varepsilon ,W,\lambda _{\mathtt{th}}^{\mathtt{off}})$ as
\begin{align}
\label{remark_uniform_th_low_1}
{h_3}({N_0},\alpha ,b,\varepsilon ,W,\lambda _{\mathtt{th}}^{\mathtt{off}}) = {A_1} \cdot {h_4}(W) + {A_2} \cdot {h_5}(W),
\end{align}
where $
{h_4}(W) = W \cdot \bigg(\frac{{{e^{({2^{\frac{b}{W}}} - 1){L_1}}} - 1}}{{{L_1}}} - \frac{{{e^{({2^{\frac{b}{W}}} - 1){L_2}}} - 1}}{{{L_2}}}\bigg)
$
and
$
{h_5}(W) = W\frac{{{e^{({2^{\frac{b}{W}}} - 1){L_2}}} - 1}}{{{L_2}}}
$
are functions of $W$, ${A_1} = \frac{{u\Gamma {N_0}\lambda _{th}^1}}{{ - \ln (1 - \varepsilon )D r_0^\alpha }}\frac{{2\pi }}{{\alpha  + 2}}(R_0^{\alpha  + 2} + \frac{{\alpha r_0^{\alpha  + 2}}}{2})$, ${A_2} = \frac{{u\Gamma {N_0}\lambda _{th}^1}}{{ - \ln (1 - \varepsilon ) D r_0^\alpha }}\pi R_s^2d_M^\alpha$, ${L_1} = \lambda _{th}^1\pi R_0^2$, and ${L_2} = \lambda _{th}^1\pi (R_0^2 - R_s^2)$ are constants. In the following, we   prove that the first-order derivatives of $h_4(W)$ and $h_5(W)$ are both negative when $W>0$.

The first-order derivative of $h_5(W)$ is $h_5^\prime(W)= \frac{{{e^{{L_2}({2^x} - 1)}}(1 - {2^x}x{L_2}\ln 2) - 1}}{{{L_2}}}$ where $x = \frac{b}{W}>0$. We further calculate the second-order derivative of $h_5(W)$ as $h_5^{\prime\prime}(x) = \frac{{e^{{L_2}({2^x} - 1)}}{2^x}x{L_2}{(\ln 2)^2}( - {L_2}{2^x} - 1)}{L_2}$. It can be seen that $h_5^{\prime\prime}(x)<0$ when $x>0$. Thus, $h_5^\prime(x)<h_5^\prime(0)=0$ when $x>0$, i.e., $h_5^\prime(W)<0$ when $W>0$. Similarly, we can prove that: $h_4^\prime(W)<0$ when $W>0$. Combining these results with \eqref{remark_uniform_th_low_1}, we can obtain that the partial derivative of $W$ is negative. Thus, ${h_3}({N_0},\alpha ,b,\varepsilon ,W,\lambda _{\mathtt{th}}^{\mathtt{off}})$ decreases with $W$, then $\lambda _{\mathtt{th}}^{\mathtt{off}}$ increases with it.
To conclude, all the properties in Corollary 1 are proved.

\section{Proof of Proposition 4}
By using a similar method as that in Appendix E, it is easy to verify the  existence and uniqueness of $\lambda _{\mathtt{th}}^{\mathtt{off}}(m)$ and $\lambda _{\mathtt{th}}^{\mathtt{on}}(m)$ via bisection searching. We now prove the case with low user density: If ${\lambda _m} < \lambda _{\mathtt{th}}^{\mathtt{off}}(m)$, regardless of the operation modes of other SBSs, we should always turn off the SBS $m$. Let the set $\mathcal{M}_0$ represent all the sleeping SBSs when SBS $m$ is turned on. Then if SBS $m$ is turned off, the  HetNet power consumption  is changed by
 \begin{small}
\begin{align}
\label{lamda_th_low_m_1}
\Delta P^{Het} \overset{(a)}{=}&\bigg(\frac{{2\pi {\lambda _0}}}{{\alpha  + 2}}(R_0^{\alpha  + 2} + \frac{{\alpha r_0^{\alpha  + 2}}}{2}) - {\lambda _0}\sum\limits_{n = 1}^M {\pi R_s^2d_n^\alpha  + \sum\limits_{n \in {M_0}} {{\lambda _n}\pi R_s^2d_n^\alpha }  + } {\lambda _m}\pi R_s^2d_m^\alpha \bigg)  \bigg(f\big(\underline L  + \sum\limits_{n \in {M_0}} {{\lambda _n}\pi R_s^2}  + {\lambda _m}\pi R_s^2\big) \nonumber\\
 & -f\big(\underline L  + \sum\limits_{n \in {M_0}} {{\lambda _n}\pi R_s^2}\big)\bigg)+ f\big(\underline L  + \sum\limits_{n \in {M_0}} {{\lambda _n}\pi R_s^2}\big)  {\lambda _m}\pi R_s^2d_m^\alpha  - \Delta p\nonumber\\
\overset{(b)}{\le} & \bigg(\frac{{2\pi {\lambda _0}}}{{\alpha  + 2}}(R_0^{\alpha  + 2} + \frac{{\alpha r_0^{\alpha  + 2}}}{2}) - {\lambda _0}\sum\limits_{n = 1}^M {\pi R_s^2d_n^\alpha  +\!\!\!\!\!\! \sum\limits_{n = 1,n \ne m}^M \!\!\!\!\!\!{{\lambda _n}\pi R_s^2d_n^\alpha }  + } {\lambda _m}\pi R_s^2d_m^\alpha \bigg)   \bigg(f\big(\underline L  + \sum\limits_{n = 1,n \ne m}^M {{\lambda _n}\pi R_s^2}  + {\lambda _m}\pi R_s^2\big) \nonumber\\
 & -f\big(\underline L  + \sum\limits_{n = 1,n \ne m}^M {{\lambda _n}\pi R_s^2}\big)\bigg)+ f\big(\underline L  + \sum\limits_{n = 1,n \ne m}^M {{\lambda _n}\pi R_s^2}\big)   {\lambda _m}\pi R_s^2d_m^\alpha  - \Delta p < 0
\end{align}
\end{small}where $(a)$ follows due to the fact that $f(x)$ in \eqref{C_fun} increases over $x \in (0, + \infty )$, and $(b)$ follows   due to \eqref{lamda_th_1_m} and the increasing of $ \lambda _{\mathtt{th}}^{\mathtt{off}}(m)$ over $[0, + \infty )$. It is then easy to verify that the HetNet's power consumption can be reduced by turning off the SBS $m$. Thus,  we should deactivate SBS $m$. Thus, the case with low user density follows. By using the similar proof method for the low user density regime, one can easily prove the case with  the  high user density, and thus is omitted here for brevity. Proposition 4 thus follows.

\bibliographystyle{IEEEtran}

\end{document}